\documentclass[aps,amsmath,amssymb,amsfonts,prd,twocolumn,groupedaddress,nofootinbib,floatfix,superscriptaddress]{revtex4-1}

\usepackage{graphicx,bm,color}
\usepackage{aas_macro}
\usepackage{multirow}
\usepackage{verbatim}
\usepackage{ulem}
\usepackage{mathtools}
\usepackage{colortbl}
\usepackage[dvipsnames]{xcolor}
\usepackage{braket}

\usepackage[pdftex,colorlinks,pdfpagelabels]{hyperref}
\hypersetup{
   bookmarksnumbered,
   citecolor={blue},
   linkcolor={blue},
   urlcolor={blue},
   filecolor={blue}
}

\definecolor{pink1}{RGB}{226, 24, 166}

\newcommand{\taon}{$\tau$-lepton }
\newcommand{\taons}{$\tau$-leptons }

\newcommand{\lsim}{\mathrel{\hbox{\rlap{\lower.75ex \hbox{$\sim$}} \kern-.3em \raise.4ex \hbox{$<$}}}}
\newcommand{\gsim}{\mathrel{\hbox{\rlap{\lower.75ex \hbox{$\sim$}} \kern-.3em \raise.4ex \hbox{$>$}}}}

\begin{document}

\title{Modeling of the Tau and Muon Neutrino-induced Optical Cherenkov Signals \\
from Upward-moving Extensive Air Showers}

\author{A.L. Cummings}
\affiliation{Gran Sasso Science Institute (GSSI), Via F. Crispi 7, 67100 L'Aquila, Italy}
\affiliation{INFN-Laboratori Nazionali del Gran Sasso, Via G. Acitelli 22, Assergi (AQ), Italy}

\author{R. Aloisio}
\affiliation{Gran Sasso Science Institute (GSSI), Via F. Crispi 7, 67100 L'Aquila, Italy}
\affiliation{INFN-Laboratori Nazionali del Gran Sasso, Via G. Acitelli 22, Assergi (AQ), Italy}

\author{J.F. Krizmanic}
\affiliation{CRESST/NASA Goddard Space Flight Center, Greenbelt, MD 20771, USA}
\affiliation{University of Maryland, Baltimore County, Baltimore, MD 21250, USA}

\date{\today}

\begin{abstract}
We present a detailed modeling and computation methodology to determine the optical Cherenkov signals produced by upward-moving extensive air showers (EASs) induced by \taons and muons, sourced from the interaction of high-energy astrophysical neutrinos interacting in the Earth. Following and extending the physics modeling and Cherenkov signal simulations performed in \cite{Reno:2019jtr}, this scheme encompasses a new, state-of-the-art computation of the muon neutrino propagation inside the Earth and the contribution to the \taon muon decay channel. The modeling takes into account all possible \taon decay and muon energy loss channels that feed the optical Cherenkov emission, produced by both tau and muon initiated EASs. The EAS modeling uses the electron energy, angular, and lateral distributions in the EAS and their evolution as well as the wavelength dependence of the Cherenkov emission and its atmospheric attenuation. The results presented here are focused on the detection capabilities of sub-orbital (balloon-borne) and orbital (satellite) based instruments.  The latter case was calculated for POEMMA\footnote{the Probe Of Extreme MultiMessenger Astrophysics:  \cite{Anchordoqui:2019omw,Olinto:2019mjh,Olinto:2017xbi}} to compare to that presented in \cite{Reno:2019jtr}, specifically including the muon-decay channel of \taons and the muonic EAS Cherenkov signal from muon neutrino interactions in the Earth. By detailing all these individual contributions to the optical Cherenkov emission and detection, we show how the ensemble that includes muonic channels provides a large detection capability for space-based, high-energy cosmic neutrino detection. Specifically, we show that for neutrino energies $\lsim$ 10 PeV, the upward-EAS sensitivity due to muon neutrino interactions in the Earth begin to dominate over that for tau neutrino interactions, effectively extending the neutrino sensitivity to lower energies. 

\end{abstract}

\maketitle

\section{Introduction}

High-energy astrophysical neutrinos and antineutrinos (hereafter collectively denoted as neutrinos) carry unique information about the most energetic non-thermal hadronic processes in the Universe and their cosmological evolution. This information provided by the neutrino channel can help to explain the physics of Ultra High Energy Cosmic Ray (UHECR) acceleration and propagation, and the UHECR nuclear composition, and to help probe new physics beyond the Standard Model (SM) \cite{Beresinsky:1969qj,Gaisser:1994yf,Ahlers:2017wkk,Ackermann:2019ows,Ackermann:2019cxh,Palladino:2020jol,Aloisio:2015ega,Yoshida:1996ie,Sigl:1998vz,Berezinsky:1999az,Aloisio:2006yi,Kotera:2010yn,AlvesBatista:2018zui,Aloisio:2015lva}. 

Being neutral and having extremely low interaction cross sections, even at the highest energies, astrophysical neutrinos are ideal astronomical messengers as they travel cosmological distances with virtually no deflection, scattering, or absorption. Moreover, in a multi-messenger approach, the observation of high-energy astrophysical neutrinos is complementary to other channels of observation such as gamma rays, cosmic rays and gravitational waves, as neutrinos bring unique information about source characteristics \cite{Ackermann:2019ows,Ahlers:2017wkk,Gaisser:1994yf}. In addition, long baseline neutrino oscillation from cosmic sources provide (to first order) a 1:1:1 ratio of neutrino flavors at Earth regardless of the flavor composition at the source \cite{Learned:1994wg}. 

The weak interaction of neutrinos requires very large amounts of target material to allow for observable interactions, which can be reached only if vast portions of the Earth and/or its atmosphere are used as targets. At neutrino energies $E_\nu \gsim 1$ PeV, the charged-current, neutral-current cross sections and ineslaticity of neutrinos and antineutrinos are essentially identical, with the cross sections increasing with neutrino energy, e.g. Ref. \cite{Block:2014kza}.
However, the expected astrophysical flux of neutrinos decreases with increasing energy, making it even more difficult to measure a statistically significant number of events at the highest energies without increasingly large detection volumes. 

The first detection of a flux of high-energy neutrinos (above the atmospheric neutrino background) coming from outside the solar system was performed by the IceCube collaboration in the energy range that spans from $\sim10$ TeV up to a few PeV \cite{Aartsen:2020aqd,Stettner:2019tok,Aartsen:2014gkd,Aartsen:2013bka,Aartsen:2013jdh}. This fundamental step forward marks the beginning of high-energy neutrino astrophysics. 

IceCube is a km$^3$ scale instrument located in the South Pole which observes high-energy neutrinos by using the Antarctic ice as a target. The Cherenkov emission induced in the ice by the products of neutrino interaction is then collected by the optical modules that instrument the cubic kilometer \cite{Williams:2020mvu}. The event topology (track-like, cascade-like, and double pulse/bang) provide a method to detect the flavor of the incident neutrinos.

Paving the way towards high-energy neutrino astronomy, in the coming age of multi messenger astrophysics, the IceCube collaboration performed several remarkable observations such as the detection of both tau and electron neutrinos fluxes \cite{Aartsen:2020aqd}, thus further confirming the cosmic origin of the signal, as well as the detection of a neutrino event with energy around $300$ TeV correlated in both arrival direction and time to a flaring blazar  observed via $\gamma$ ray emission by the FermiLAT satellite \cite{IceCube:2018dnn}. However, given its experimental acceptance, IceCube currently has limited sensitivity at energies larger than a few PeV \cite{Ahlers:2017wkk}. Other neutrino observatories of km$^3$ scale, based on the Cherenkov light detection in water and still under construction in undersea environments, like the KM3NeT \cite{Adrian-Martinez:2016fdl} and GVD detectors \cite{Avrorin:2019dov}, share similar limits in sensitivity and angular resolution compared to IceCube. 

Notwithstanding the transformational results obtained so far, to go beyond the discovery phase, as commonly happens to astronomical instruments, an improvement in exposure and pointing resolution of the detectors is required. In order to maximize the experimental exposure, i.e. the target material, the techniques proposed to detect astrophysical neutrinos are spread over different environments: underground, ground, sub-orbital or orbital. Underground experiments, such as IceCube or KM3NeT and GVD, focus on the upward-going Cherenkov emission produced in transparent underground environments, such as ice or water, by the products of charged-and neutral-current interactions of neutrinos (see \cite{Palladino:2020jol} and references therein). 

At energies larger than few PeV, in order to maximise the detector acceptance, experiments can also use the coherent radio emission produced by the time-varying net charge in neutrino initiated showers in dense media such as ice, the so-called Askaryan effect \cite{Askaryan:1962aa,Askaryan:1989hk}. The balloon-borne ANITA experiment \cite{Barwick:2005hn,2009APh....32...10A}, pioneered the principle of radio neutrino detection, based on the Askaryan effect, set definitive neutrino flux limits at the highest enegies ($E_\nu \gsim 100 EeV$) \cite{Gorham:2019guw}. Radio waves attenuate in ice with a path length at the km scale, thus a limited number of ground-based radio antennas can cover large volumes easily reaching one hundred of cubic kilometres. In the last decade, two experiments, both performed in Antarctica, have tested the Askaryan effect to detect astrophysical neutrinos: the ARA detector on the South Pole \cite{Allison:2011wk,Allison:2015eky}  and the ARIANNA detector on the Ross Ice Shelf  \cite{Barwick:2014rca,Barwick:2014pca}. These experimental efforts should comply with a suitable reduction of the strong background of thermal and anthropogenic radio signals, the observations conducted so far have enabled the determination of upper limits to the astrophysical neutrino flux at energies larger than 10 PeV \cite{Allison:2019xtn}. The detection of radio signals are also a significant part of the proposed IceCubeGenII experiment \cite{Aartsen:2019swn}, an extension of the actual IceCube infrastructure that, together with a factor of 10 increase in the target volume for Cherenkov light detection in ice, promises a larger acceptance at the highest energies.

Another class of experiments focus on detecting the optical and radio signals from neutrino-induced extensive air showers (EASs). Astrophysical neutrino EAS detection in the case of ground, sub-orbital and space-based  experiments use the high-energy electrons, muons,  and/or taus coming from neutrino charged-current interaction either in the Earth or in the Earth's atmosphere as the seeds of the EAS \cite{Halzen:1998be,Fargion:2000iz,Feng:2001ue,Bertou:2001vm,Bottai:2002nn,Fargion:2003kn,Tseng:2003pn,Aramo:2004pr,Dutta:2005yt,Asaoka:2012em,Otte:2018uxj,Aab:2019auo,Adams:2017fjh,Wiencke:2019vke,Anchordoqui:2019omw,Olinto:2017xbi,Aab:2019ogu,Aab:2019auo}. 

Ground-based experiments designed to detect UHECR, such as the Pierre Auger Observatory \cite{Aab:2019ogu} and the Telescope Array \cite{AbuZayyad:2012kk,Tokuno:2012mi} are able to observe EASs produced in the atmosphere through the (isotropic) fluorescence emission and from direct detection of charged particles on ground \cite{Aloisio:2017ooo}. The analysis of the data collected by these experiments have shown no direct evidence for neutrino events, providing upper limits to the possible astrophysical neutrino flux at energy larger than $0.1$~EeV \cite{Aab:2019auo}.

Given the large amount of atmosphere that can be observed by sub-orbital and space-based experiments, they can employ larger target masses with respect to ground and underground experiments, thus providing significant gains in the overall acceptance. The detection of EAS in the atmosphere from sub-orbital and space-based detectors can be achieved through both: i) the beamed Cherenkov light produced in the atmosphere by charged particles from the EAS (mainly electrons) and ii) the coherent radio emission produced by the Askaryan (negative charge excess of the EAS) and the geomagnetic (charge separation induced in the EAS by the geomagnetic field) effects at lower energies. At higher energies, the fluorescence detection of EAS induced from neutrino interactions in the atmosphere can be used to achieve extraordinary sensitivity \cite{Anchordoqui:2019omw}.  Both the optical and radio detection techniques provide good resolution in the determination of the primary neutrino energy and have an excellent angular resolution in the reconstruction of the incident neutrino direction. 

Coherent radio emission produced by UHECR induced EASs has been observed by the ANITA balloon-borne experiment \cite{Schoorlemmer:2015afa}, through four long-duration balloon flights. The ANITA detector has demonstrated the viability of the radio detection technique on downward (reflected) and horizontal (direct) EASs produced by UHECRs \cite{Hoover:2010qt,Schoorlemmer:2015afa}. However, the unique topology of a neutrino-induced event emerging from the Earth associated with the upward direction of an EAS, and a search for these signals by ANITA has led to upper limits in the astrophysical neutrino flux at energies larger than 10 PeV \cite{Barwick:2005hn,Allison:2018cxu,Gorham:2019guw}. Recently, the ANITA collaboration has announced the detection of a few events consistent with upward going EASs \cite{Gorham:2018ydl}. However, given the inferred path-length through Earth and the reconstructed energy of the signal, it is unlikely that these events can be ascribed to high-energy astrophysical neutrinos \cite{Gorham:2016zah,Gorham:2018ydl,Romero-Wolf:2018zxt,Cummings:2019pkx}. 

In the present paper we will focus on modeling the optical Cherenkov emission of the EASs produced in the atmosphere by the decay or interaction of leptons resulting from the interaction of high-energy neutrinos inside the Earth. If a neutrino interaction occurs close enough to the Earth's surface, a lepton (muon or tau) can emerge into the atmosphere and produce an EAS, the so-called Earth-skimming neutrino events, although we will show that the Earth emergence angles can be fairly large at the lower neutrino energies ($E_\nu \lsim 100$ PeV). We will concentrate our analysis on the detection capabilities of space-based and sub-orbital detectors to these events. 

Several different proposals were recently presented to detect Earth-emergent neutrinos through the optical Cherenkov emission from an EAS in the form of ground-based experiments (Trinity \cite{Otte:2018uxj}), 
sub-orbital (EUSO-SPB2 \cite{Adams:2017fjh,Wiencke:2019vke}) and orbital experiments (POEMMA \cite{Anchordoqui:2019omw,Olinto:2019mjh,Olinto:2017xbi}). The observation of these signals enables, in principle, huge neutrino target masses with increased acceptance at the highest energies with respect to underground experiments, by using the Earth's atmosphere as part of the `detector',  even with a reduced duty-cycle (at the level of $\sim 20\%$ compared to $\sim100\%$ of the underground approach).

As neutrinos propagate through the Earth, they may undergo neutral-current interactions which reduce the initial energy or charged-current interactions, which convert the neutrino into its corresponding lepton. As leptons propagate through the Earth, they may undergo severe energy losses (with electrons losing the most, followed by muons and $\tau$-leptons, roughly in the ratio of their respective masses) and also, excluding electrons, have the potential to decay back into a neutrino (the so-called regeneration effect). Implementing a new computation scheme to study neutrino propagation inside the Earth \cite{Alvarez-Muniz:2017mpk}, we will present results that are an extension of the approach already discussed in \cite{Reno:2019jtr}, where the authors considered only the hadronic and electronic decay channels of \taons. Here we extend the computation to include the case of muons produced by \taon decay and by muon neutrino interaction inside the Earth. In these two cases, the interacting muon will be the EAS initiating particle, i.e. a high-energy muonic particle cascade.

The paper is organized as follows: in section \ref{earth} we discuss the interaction of neutrinos inside the Earth, in section \ref{upward} we review the physics of upward EAS, taking into account both muon and \taon initiated EAS, together with a discussion on the expected Cherenkov emission, in section \ref{rates} we compute the event rate expected for space-based and sub-orbital experiments specifying the discussion to the cases of POEMMA (525 km altitude) and EUSO-SPB2 (33 km altitude), conclusions take place in section \ref{conclusions}.

\section{Neutrino Interactions in Earth}
\label{earth}

The detection of astrophysical neutrinos through an Earth target requires an accurate description of neutrino propagation inside the Earth. At high energies, the Earth becomes opaque to neutrinos and thus the detection signal is provided by `Earth-skimming' events, that traverse a relatively short slant depth in the Earth \cite{Gandhi:1995mpk}. Thus, a detailed model of the Earth density distribution along the neutrino trajectories is required as determined by the detection geometry for the muon and \taon initiated EASs.

In this paper, we have used the computation scheme, based on a Monte Carlo approach, called NuTauSim\footnote{One of the authors of the present work (A.L. Cummings) contributed to the development of the NuTauSim computation scheme \cite{Alvarez-Muniz:2017mpk}.} \cite{Alvarez-Muniz:2017mpk}, which propagates tau neutrinos inside the Earth using the Preliminary Reference Earth Model (PREM) \cite{Dziewonski:1981xy} taking into account both tau neutrino charged-current and neutral current interactions, \taon energy losses and tau neutrino regeneration effects. The NuTauSim code determines the output flux and energy spectra of tau neutrinos and \taons given an input flux of tau neutrinos, neutrino energy (or spectra) and particle Earth-emergence angle. The variable inputs for this kind of computation are 1) Neutrino spectra 2) Ice layer thickness of the Earth 3) Model of \taon energy losses at high energies 4) Model of neutrino neutral-current and charged-current cross sections at high energies. A complete description of the NuTauSim code and its primary results are given in \cite{Alvarez-Muniz:2017mpk}. In all computations that follow, we have assumed the ice layer around the Earth equal to 4~km, compared to the 3 km thickness in the PREM.

The main source of uncertainties in the determination of the \taon fluxes comes from the uncertainties in the neutrino-nucleon cross section and the \taon energy losses at the highest energies, as they require the extrapolation of the nucleon structure functions at energies not probed experimentally.

In the case of the neutrino-nucleon cross section, the standard model uncertainties can reach the level of a factor of five at the highest energies (up to $10^{21}$ eV) depending on the extrapolations of the nucleon structure function \cite{Connolly:2011vc}. In the NuTauSim approach, the results presented for the neutrino-nucleon cross section in \cite{Connolly:2011vc} are used, which include an upper, lower, and middle model. For what follows, we use the middle extrapolated model. Other approaches based, for instance, on the dipole model \cite{Jeong:2017mzv}, provide a slightly lower estimation of the neutrino-nucleon cross section of roughly 20$\%$ at energies larger than $10^{18}$ eV with closer agreement at lower energies.

\begin{figure}[t]
\includegraphics[width=\linewidth]{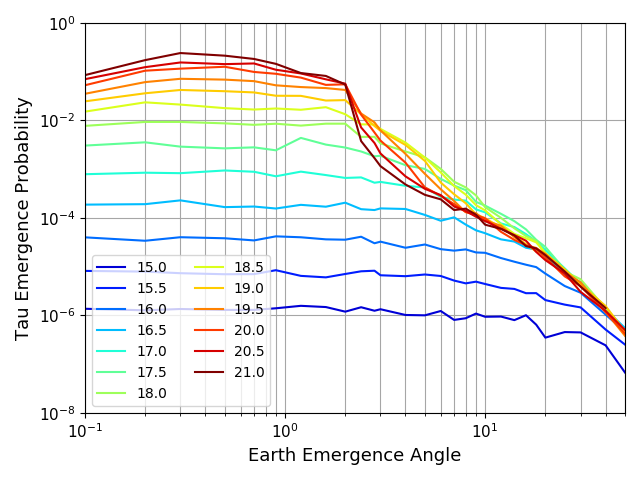}
\includegraphics[width=\linewidth]{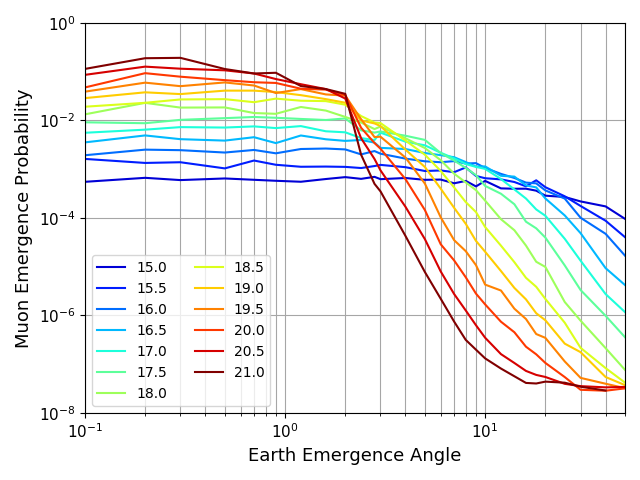}
\caption{[Upper panel] Tau emergence probability as a function of Earth emergence angle (in degrees) for energies ranging from $10^{15}$~eV to $10^{21}$~eV as labeled. [Lower panel] Muon emergence probability for the same parameter space above.}
\label{Emergence_Probabilities}
\end{figure}

The model used in NuTauSim to describe the \taon energy losses takes into account all relevant channels: ionisation, pair production, bremsstrahlung and photo-nuclear interactions, being the latter dominant above 1~PeV. The NuTauSim computation scheme is based on either the parameterization of \taon energy losses given by Armesto, Salgado and Wiedemann (ASW) \cite{Armesto:2004hh} or Abramowicz,  Levin, Levy and Maor (ALLM)\cite{Abramowicz:1991xz,Abramowicz:1997ms,Dutta:2000hh}. For what follows, we use the ALLM approach that uses the continuous model for energy losses: $-\frac{dE}{dX} = a(E) + b(E)E$ where $a(E)$ and $b(E)$ are parameterized for the \taon in \cite{Dutta:2000hh}. In the simulation, for the propagation of the \taon inside the Earth, a cutoff energy of 0.1~PeV is assumed, noting that this alters the Earth emergence probability by less than $0.1\%$ even for neutrinos with initial energy 1~PeV out to Earth emergence angles larger than $30^{\circ}$ (for higher energy neutrinos, this fraction is even smaller. See figure \ref{Emergence_Probabilities}).

\begin{figure}[t]
\includegraphics[width=\linewidth]{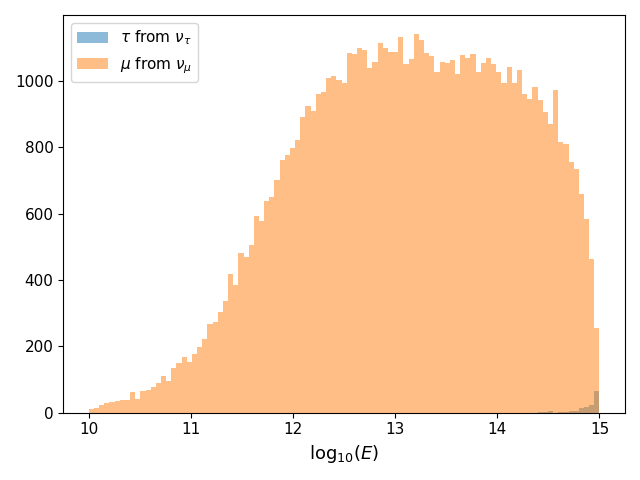}
\includegraphics[width=\linewidth]{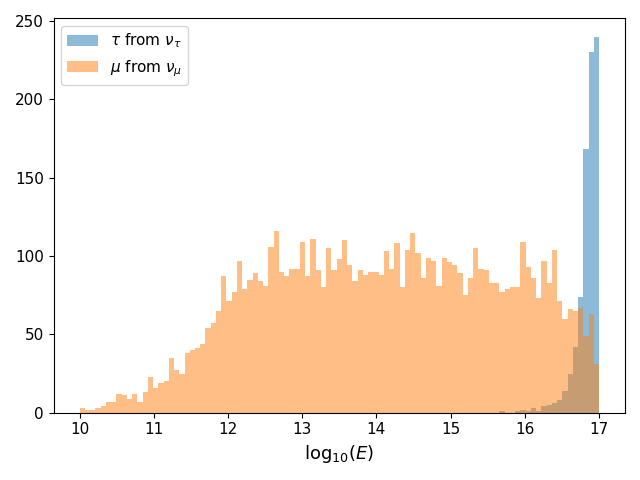}
\caption{[Upper panel] Energy distribution of emergent \taons from $\nu_{\tau}$ (blue histograms) and muons from $\nu_{\mu}$ (orange histograms)  produced by a mono-energetic flux of $10^{8}$ neutrinos with energy $10^{15}$~eV and an Earth emergence angle of $1^{\circ}$. [Lower panel] The same as in upper panel with a flux of $10^{6}$ neutrinos with $10^{17}$~eV energy. Note the small number of tau leptons in the 1~PeV distribution.}
\label{Energy_Dists}
\end{figure}

As discussed in the Introduction, we also model the EAS signal produced by muons sourced from muon neutrinos interacting in the Earth. The NuTauSim computation scheme was originally developed for tau neutrino interactions. To account for the muon neutrino propagation inside the Earth we have modified the NuTauSim code according to the following prescriptions \cite{Alvarez-Muniz:2021mpk}:

\begin{itemize}
    \item{Substitute the \taon mass and decay distance with those for the muon.}
    \item{$E_{\nu_{\mu}}$ from muon decay sampling table created using 10,000 relativistic muon decays in Pythia8 \cite{Sjostrand:2014zea} for possible $\mu \rightarrow \nu_{\mu} \rightarrow \mu$ regeneration effects.}
    \item{Average energy losses modelled as continuous with $a(E)$ and $b(E)$ parameterized in the ALLM approach as in \cite{Dutta:2000hh,Groom:2001kq} for high-energy muons.}
    \item{Low energy cutoff set to $10^{10}$~eV instead of $10^{14}$~eV to capture the complete distribution of emerging muons (see figure \ref{Energy_Dists}).}
    \item{Above $10^{15}$~eV, the parameterization for the neutrino cross section was not changed, as it is the same for $\nu_{\tau}$ as $\nu_{\mu}$. For energies below, we use the results of \cite{Gandhi:1995mpk} to model the difference between neutrinos and anti-neutrinos.}
    \item{High Lorentz factors of the resulting leptons ensure the sampling of muon energy from the parent neutrino is the same as that of the $\tau$-lepton, which is well parameterized in NuTauSim for neutrinos and anti-neutrinos for the energies of interest.}
\end{itemize}

In both cases of tau and muon neutrinos, we simulate mono-energetic fluxes from $E_\nu=10^{15}$~eV to $E_\nu=10^{21}$~eV, spaced in half decades and Earth-emergence angles (complement of $\theta_{tr}$, the trajectory angle from local Earth normal) from $0.1^{\circ}$ to $50^{\circ}$, spaced non-uniformly in 30 bins to attempt to capture the characteristic features of the curves. For high input energies and Earth emergence angles, computation time (for reasonable statistics) grows exceedingly large due to increased energy losses and number of interactions. To compensate for this, for each neutrino energy and emergence angle, we propagate neutrinos until we reach 100 emerging leptons, regardless of emergent energy, resulting in a $10\%$ statistical error on the Earth emergence probability. The energy distributions of the emerging leptons are later fit using 3-dimensional kernel density estimation to provide smoother sampling, which reproduces the energy distributions with far more events extremely well. The Earth emergence probability as a function of emergence angle for primary neutrino energies ranging from $10^{15}$~eV up to $10^{21}$~eV is shown in figure \ref{Emergence_Probabilities} for both \taons and muons.

For energies above a few TeV, the dominant energy loss term comes from the $b(E)$ factor in the continuous energy loss model. At 100~TeV, $b(E)$ is roughly $4.5\times 10^{-6} \, \mathrm{g}^{-1}\mathrm{cm}^{2}$ for muons and $3\times 10^{-7} \, \mathrm{g}^{-1}\mathrm{cm}^{2}$ for tau leptons. The muon decay length is $6.25 \times 10^{5}$~km and, assuming an average rock density of $2.9\, \mathrm{g}\,\mathrm{cm}^{-2}$, the interaction scale is calculated as $0.7$~km. For a \taon of comparable energy, the equivalent values are $4.9$~m and $11.5$~km. Except at very high energies, the \taon mean free path is dominated by the decay distance, whereas that of the muon is dictated by energy losses. Because the muon lifetime is orders of magnitude larger than that of the $\tau$-lepton, the \taon Earth emergence probability is lower than that of the muon until energies around $10^{18}$ eV, with large difference at lower energies (where energy losses become less significant):  at PeV energies the \taon emergency probability is a factor 600 below that of muons, as seen in figure \ref{Emergence_Probabilities}. 

For energies larger than $10^{18}$ eV and small Earth emergence angles, muons and \taons have more or less equal emergence probabilities, where the dominant scale becomes the neutrino interaction length. For high energies and large emergence angles, we might expect the same behaviour, but we must account for neutrino regeneration $\nu_{\ell} \rightarrow \ell \rightarrow \nu_{\ell}$. This occurs when the energy of a $\nu_{\ell}$ from a lepton decay is high enough to enable a re-interaction deeper in the Earth. For $\tau$-leptons, this is a significant effect, because a high-energy \taon has a strong chance of decaying due to its short lifetime. However, the effect is negligible for muons, as the energy losses are so significant and the decay time so long, that a $\nu_{\mu}$ from a muon decay has low energy and a negligible chance of re-interaction.

For the same reasoning, we expect to see the energy distribution of emerging muons to be more strongly spread than $\tau$-leptons, given an identical input of neutrino fluxes. In figure \ref{Energy_Dists}, we plot the energy distributions of emerging \taons and muons with Earth emergence angle of $1^{\circ}$, to demonstrate both this behaviour and the effect of differing emergence probabilities. In the upper panel of  figure \ref{Energy_Dists}, we simulate $10^8$ neutrinos with 1~PeV energy, and in the lower panel we simulate $10^6$ neutrinos with 100~PeV energy. The energy distribution of the emerging muons is fairly uniformly spread from the primary neutrino energy down to $10^{12}$~eV, where muons begin to rapidly lose energy and decay, regardless of the primary $\nu_{\mu}$ energy or emergence angle. From the lower panel of figure \ref{Energy_Dists}, it follows that, due to the amplified emergence probability, there may still be a comparable number of high-energy muons emerging from the Earth compared to $\tau$-leptons, even accounting for the increased energy losses. As a subdominant process, we also expect a muon flux from \taon decays inside the Earth. However, we ignore this effect in the results presented in the present paper.

\section{Upward Extensive Air Shower Modeling}
\label{upward}

\subsection{Tau-Lepton Contribution}
\taons have low ionization and radiative energy losses and a short lifetime compared to muons, thus \taon initiated EASs occur virtually completely via \taon decay, rather than interaction in the atmosphere. \taons are the only leptons massive enough to decay into hadrons via the weak interaction, which gives rise to a rich decay phenomenology with many channels \cite{Olive2014aaa}. We do not need to list here all the decay channels, but we may classify them into three distinct branches and list their respective probabilities:

\begin{itemize}
    \item{$\tau^{\mp} \rightarrow \mathrm{hadrons} + \nu_{\tau}(\bar{\nu}_\tau) \approx 64.79\%$}
    \item{$\tau^{\mp} \rightarrow e^{\mp} + \bar{\nu}_e(\nu_e) + \nu_{\tau}(\bar{\nu}_\tau) \approx 17.82\%$}
    \item{$\tau^{\mp} \rightarrow \mu^{\mp} + \bar{\nu}_\mu(\nu_\mu) + \nu_{\tau}(\bar{\nu}_\tau) \approx 17.39\%$}
\end{itemize}

\begin{figure}[t!]
\includegraphics[width=\linewidth]{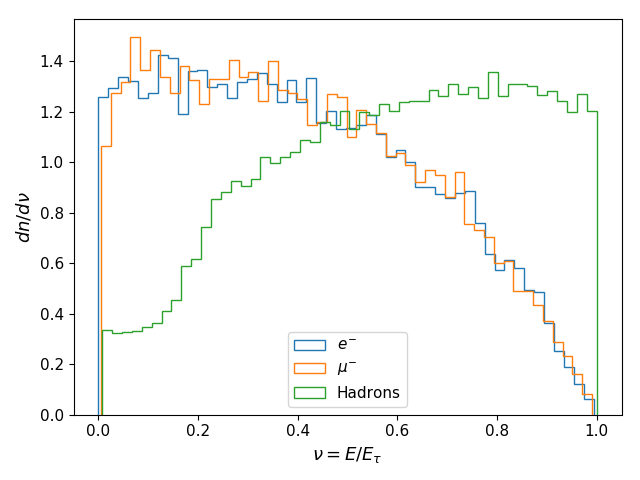}
\caption{Fractional energy distribution of a relativistic \taon decay with negative polarization. The hadronic channel is the sum of the fractional energies of all hadrons in a given decay. These values are calculated using 100,000 PYTHIA \taon decays \cite{Sjostrand:2014zea}.}
\label{Tau_Lepton_Dist}
\end{figure}

Additionally, a large fraction of the primary \taon energy is distributed to the decay products. Therefore, the high-energy decay products may produce conventional EASs after propagation through the Earth. The fractional energy distribution of a relativistic \taon decay is shown in figure \ref{Tau_Lepton_Dist}. All sampling is done on \taons with negative polarization to account for the production by the parent neutrino. This distribution can also be used to sample the decay products from a positively polarized anti-$\tau$-lepton, coming from parent anti-neutrinos because anti particles of the opposite polarization have the same decay spectrum. The average fractional energy carried by hadrons is $58\%$ of the primary \taon energy while $42\%$ is the average fractional energy that goes into leptons (muons and electrons) via the 3-body decays. At EeV-scale and above energies, the tau can decay deep in the atmosphere, with the average decay length given by $4.9\mathrm{km}(E_{\tau}/10^{17}\mathrm{eV})$.

\subsection{Muon Contribution}

There are two ways that muons with energies above 100~TeV can be produced and begin propagating in the atmosphere. Either a muon neutrino interacts close enough to the Earth surface to emerge as a muon or the muon is produced by the decay of a \taon resulting from tau neutrino interaction in the Earth, the latter case occurring with the branching ratio listed above. Here, we distinguish these two cases as primary and secondary muons respectively.

Muons have largely been neglected in calculations of upward-going EASs due to their relatively low interaction cross section and ionization yields in air \cite{Stanev1989mpk}. However, much of the work done on this subject looks only at the average losses for the muon which excludes the possibility of large one-time energy depositions in the atmosphere along the muon track. Similarly, the case of muons from muon neutrino interaction inside the Earth usually is not considered under the assumption that the large energy losses in Earth (compared to the $\tau$-lepton) will result in too low energy muons to provide a sufficiently bright EAS while also restricting the volume of the target for $\nu_\mu$ interactions. That is to say, in general, muon interactions in air have not been treated as an important process in air shower physics, as the average muon energy is not high enough to trigger any substantial sub-showers that may rival the primary electromagnetic cascade of the same energy. 

\begin{figure}[t]
\includegraphics[width=\linewidth]{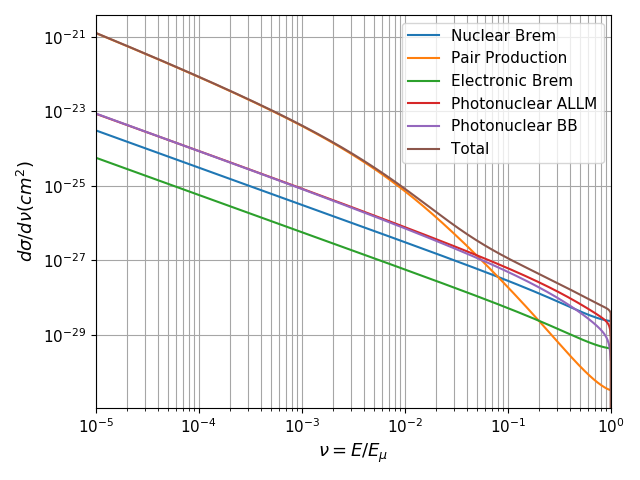}
\caption{Muon differential cross sections for nuclear and electronic bremsstrahlung \cite{Groom:2001kq}, electron-positron pair production \cite{Dutta:2000hh}, and photonuclear interactions (we show the two primary models as described by \cite{Dutta:2000hh,Bezrukov:1981ci}) as a function of fractional energy deposition for a 100~PeV muon in air.}
\label{Muon_Cross_Sections}
\end{figure}

During propagation in the atmosphere, a high-energy muon will undergo ionization and radiative energy losses. The radiative losses are capable of depositing a substantial fraction of the muon energy into a single interaction, which may trigger a conventional, electromagnetic particle cascade. We consider three main processes for muon interaction in the atmosphere: bremsstrahlung emission, photonuclear interactions, and electron-positron pair production. Bremsstrahlung interactions of muons occur on both nuclei and electrons of air molecules, whereby a high-energy gamma is radiated. Photonuclear interactions occur when a muon interacts with a nucleus or nucleon, through a virtual photon exchange  with large four momentum transferred (deep inelastic scattering), resulting in the release of high-energy hadrons. Pair production occurs when a virtual photon emitted by the muon in the external field of a nucleus is energetic enough to spontaneously generate highenergy electron-positron pairs. We ignore the possibility of muon decay, as the decay length of a muon is $6.25\times 10^{6} \Big(\frac{E}{1~\mathrm{PeV}} \Big)$~km, compared to $\mathcal{O}(10^{3}~\mathrm{km})$ of the maximum path length through the Earth's atmosphere from ground. The differential muon cross sections we consider are shown in figure \ref{Muon_Cross_Sections} and taken from \cite{Groom:2001kq,Dutta:2000hh,Bezrukov:1981ci}.

\begin{figure}[t]
\includegraphics[width=\linewidth]{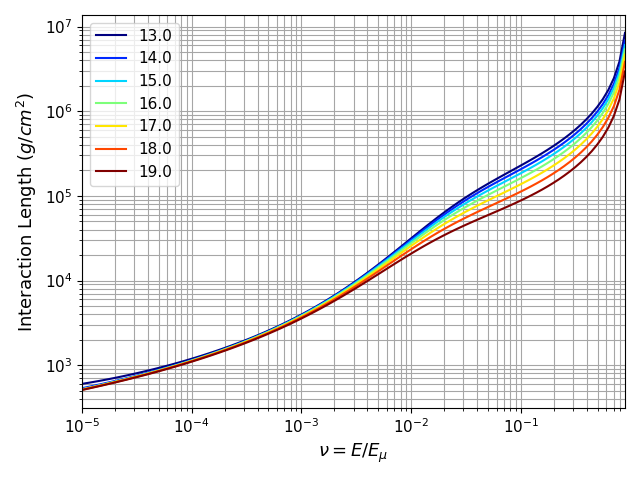}
\caption{Muon-air interaction lengths as a function of the fractional energy deposition $v=E/E_\mu$ for various muon energies as labelled. Note that a full atmosphere ranges from $1030 \frac{\mathrm{g}}{\mathrm{cm}^{2}}$ (perfectly vertical) to $34000 \frac{\mathrm{g}}{\mathrm{cm}^{2}}$ (perfectly horizontal).}
\label{Muon_Interaction_Lengths}
\end{figure}

\begin{figure}[t]
\includegraphics[width=\linewidth]{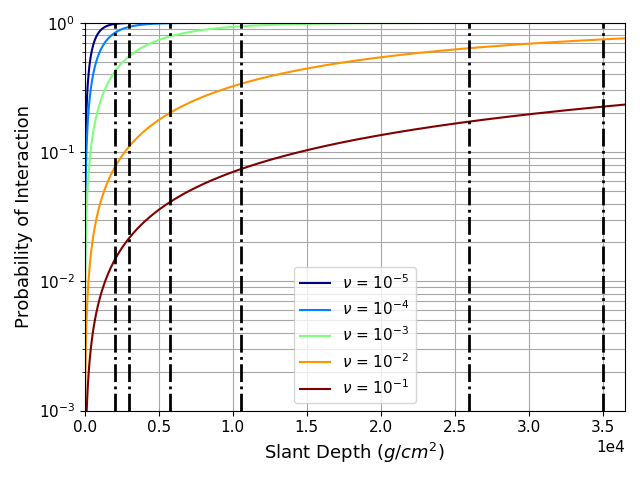}
\caption{Cumulative 100~PeV muon-air interaction probability as a function of atmospheric depth for various fractional energy depositions. Dashed lines correspond to maximum depth of the Earth atmosphere for emergence angles $30^{\circ}$, $20^{\circ}$, $10^{\circ}$, $5^{\circ}$, $1^{\circ}$, and $0^{\circ}$ (perfectly horizontal).}
\label{Muon_Probabilities}
\end{figure}

To get a sense of scale for these processes, we calculate the interaction length as a function of the fractional energy deposition, given by:

\begin{equation}
X_{\mathrm{int}}^{\mu} = \frac{1}{N \int_{v}^{1} \frac{d \sigma}{d v}}
\label{eq:Xint}
\end{equation}
where $v=E/E_\mu$ and $N$ is the number of targets in one gram of air. From equation \ref{eq:Xint}, it is simple to calculate the cumulative muon interaction probability as a function of the atmospheric slant depth $X$, given by $\mathrm{P}_{\mathrm{int}} (X) = 1-e^{-X/X_{\mathrm{int}}}$. The muon interaction length as a function of the fractional energy and the cumulative interaction probability for 100~PeV muons as a function of the atmospheric slant depth are shown in figures \ref{Muon_Interaction_Lengths} and \ref{Muon_Probabilities} respectively.

Figure \ref{Muon_Probabilities} shows that roughly $22\%$ of 100~PeV muons may begin an electromagnetic particle cascade with 10~PeV energy or larger inside a full (Earth emergence angle $0^{\circ}$) atmosphere. This is consistent with Monte Carlo simulations of muon induced air showers we have performed with CORSIKA (see next section) \cite{Heck:1998vt} \cite{Cummings:2019rlt}. This result indicates that high-energy muons, whether they come from muon or tau neutrino interactions in the Earth, have a non-negligible chance to interact in the atmosphere and initiate conventional electromagnetic EASs of significant brightness. 

On average, particle cascades induced by muon interactions start fairly deep in the atmosphere due to the low interaction cross sections. Indeed, given the long muon decay paths at high-energy, muons can initiate electromagnetic sub-showers along their entire trajectory, initiating high-altitude EAS's such that a sub-orbital instrument may be situated inside the development of the particle cascade itself. Concerning the Cherenkov emission of the cascade, the high-altitude EAS leads to strongly reduced atmospheric attenuation and a more focused Cherenkov cone. This also implies that the near-field effects of viewing the optical Cherenkov signal from close by EAS can be important. Moreover, the Cherenkov emission intensity from muon induced showers can mimic the brightness of much higher energy tau lepton induced showers occurring lower in atmosphere. For these reasons, we consider muons as an additional channel of Cherenkov light generation. 

Finally, we do not consider electron neutrino interaction in the Earth as an upward-moving EAS signal source, because the produced electrons suffer high-energy losses inside the Earth being practically completely absorbed.

\subsection{Particle Cascades and Extensive Air Showers}

In order to describe the longitudinal development of EAS in the atmosphere, we have chosen to use a full Monte Carlo simulation and compared the results to a phenomenological parameterization. We have simulated particle profiles in the EAS using a slightly modified version of the CORSIKA-75600 \cite{Heck:1998vt} computation scheme, which was properly modified to account for upward moving showers \cite{ThanksDieter}. Our CORSIKA executable was compiled with the hadronic interaction model QGSJETII-04 \cite{Ostapchenko:2010vb,Ostapchenko:2013pia}  at high-energy, i.e. larger than 80~GeV in the lab frame, and the GHEISHA 2002d \cite{Fesefeldt:1985yw} low energy interaction model, at energies below 80~GeV in the lab frame. The computation scheme we used encompasses the options for thinning (with thinning ratio $\epsilon = 10^{-6}$, and maximum weighting factor $W=100$ for primary energies above $10^{17}$~eV, and no thinning below), slant depth (such that the longitudinal profile is computed as a function of the total depth through the atmosphere, and not as a function of altitude), curved atmosphere and upward geometry of the showers. 

The model of atmosphere used in our computations, embedded in the CORSIKA code, is the 1976 US standard atmosphere \cite{USatmo:1976aaa}. The longitudinal profiles as a function of slant depth are generated starting at ground with $0^{\circ}$ Earth-emergence angle (perfectly horizontal) and later shifted according to the atmospheric depth profile taking into account the particle trajectory angle and the decay length for \taons (interaction length for muons). This assumes that longitudinal development of upward showers is independent of the shower trajectory angle and starting position in the atmosphere. To a large degree, this is a fair assumption, with the main difference resulting from the competing processes of decay and interaction of the high-energy hadrons which fuel the shower development. The decay length of a charged pion is given by $56~\mathrm{km}(E_{\pi}/\mathrm{TeV})$, to be compared with the interaction length in air that diminish with increasing pion energy, being around $108 \, \mathrm{g} \, \mathrm{cm}^{-2}$ for 1~TeV pions \cite{Gaisser:2016uoy}. At sea level, the corresponding interaction length is 0.87~km, increasing to 59.79~km at 30~km altitude (the highest starting altitude we simulate). This indicates that the longitudinal shower development is dominated by pion interactions and our approximation is valid as the pion energy exceeds roughly 1~TeV, which is a condition extensively realized in our simulations.

\begin{figure}[t!]
\includegraphics[width=\linewidth]{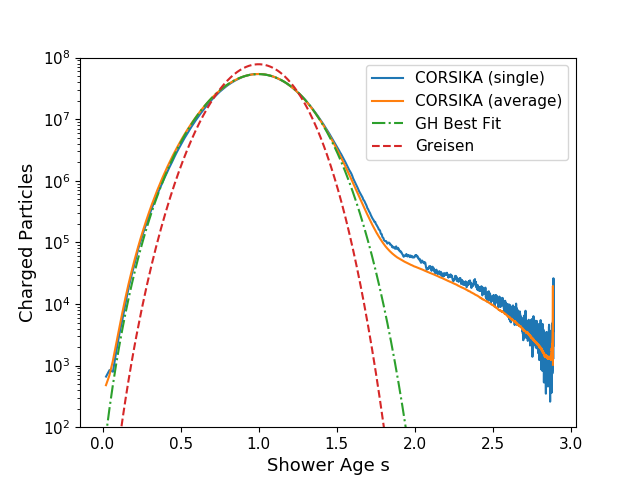}
\caption{Electron-Positron longitudinal profile as a function of shower age for an upwards 100~PeV proton shower as simulated in CORSIKA and fit with appropriate Gaisser-Hillas parameters (assuming fixed $\lambda$) compared to the Greisen parameterization for an electromagnetic shower \cite{Gaisser:2016uoy}.}
\label{GHComp}
\end{figure}

Using a Monte Carlo approach rather than a shower parameterization has several advantages for modeling the longitudinal development of upward-moving EASs. First, to our knowledge, all common shower parameterizations were developed to describe downward-going EAS's. While describing the longitudinal profiles in terms of slant depth should be fairly accurate in the description of upward-moving EASs, there could be differences especially late in shower ages due to the difference in upgoing versus downgoing trajectories in an exponential atmosphere, especially in terms of the hadronic interaction/decay argument discussed above. The modeling of the particle content at late shower ages (large atmospheric depths) can strongly influence the Cherenkov generation and may have increased geometric acceptance due to observing the EAS from close range. A comparison between 100~PeV proton showers simulated in CORSIKA and the corresponding Gaisser-Hillas fit and the Greisen EM shower parameterization \cite{Gaisser:2016uoy} in figure \ref{GHComp} shows strong disagreement at late shower ages.

We have generated EAS longitudinal particles profiles for $\tau$-lepton, proton, and gamma initiated showers at energies ranging from 1~PeV to 10~EeV to study the shower properties of each primary and quantify the approximation of a \taon induced shower as a proton induced shower of comparable energy. When a Gaisser-Hillas profile fit is made to both the longitudinal profiles of \taon induced showers and proton induced showers (with primary energy sampled from PYTHIA decays), the profile parameters are in quite good agreement, justifying this approximation \cite{Cummings:2019rlt}. 

Concerning the phenomenology associated to muon-initiated EASs, we generated muon EAS profiles in the same energy range as \taons to check our analytic approach for muon energy losses in air, and found good agreement \cite{Cummings:2019rlt}. However, for the main work presented here, we used the CORSIKA Monte Carlo code to simulate 1000 longitudinal profiles generated by 100~PeV proton air showers and appropriately scale the average of these showers to represent the EASs produced via \taon or muons. This captures the average behavior of a proton induced EAS well, but notably lacks inherent shower to shower fluctuations in starting point, which can become important at high altitudes, and deserves a future study. For muon induced showers, it could be argued that the longitudinal profiles of electron or gamma induced showers should be used (which can yield a $\sim 25\%$ increase in charged particle content than proton induced showers). However, the dominant channel at high muon energies and high fractional energy deposition (see figure \ref{Muon_Cross_Sections}) is through photo-nuclear interactions. For this reason, the final state which forms the EAS for both \taon and muon initiated showers both include hadrons. Therefore, the approximation of using a proton-initiated EAS still provides a better description and we took a conservative approach always using this approximation. 

\subsection{Atmospheric Model}
The model we use for air density as a function of altitude is the same as that used in the CORSIKA simulation \cite{Heck:1998vt} \cite{USatmo:1976aaa}, which is described by:

\begin{equation}
\rho(z) = 
\begin{cases}
    \frac{b}{c} e^{-z/c}& \text{if } z\leq 100 ~\mathrm{km}\\
    \frac{b}{c}              & \text{if } z > 100 ~\mathrm{km}
\end{cases}
\end{equation}

with parameters $b$ and $c$ for US Standard Atmosphere given by: 

\begin{center}
\begin{tabular}{ |c |c |c |}
\hline
 $z (km)$ & $b (g/cm^{2})$ & $c (cm)$ \\
 \hline
 0-4 & 1222.6562 & 994186.38 \\  
 4-10 & 1144.9069 & 878153.55 \\
 10-40 & 1305.5948 & 636143.04 \\
 40-100 & 540.1778 & 772170.16 \\
 100-112 & 1 & $1 \times 10^{9}$ \\
 \hline
\end{tabular}
\end{center}

Slant depth is calculated via:

\begin{equation}
\begin{split}
X &= \int^{z}_{0} \rho(z) dl\\
dl &= \frac{z+R_{E}}{\sqrt{R_{E}^{2}\mathrm{cos}^{2}\theta_{tr} +z^{2}+2 z R_{E}}} dz
\end{split}
\end{equation}

Where $dl$ is the path length through the curved atmosphere, $R_{E}$ is the spherical Earth's radius, and $\theta_{tr}$ is the trajectory angle 
(respect to the Earth normal or $\frac{\pi}{2}$ minus the Earth emergence angle, see Fig.~\ref{Geometry}). An analytical integral is difficult to evaluate, so an interpolation table between $z$, $l$, and $X$, is calculated numerically with steps in $X$ of 0.01~$\mathrm{g}/\mathrm{cm}^{2}$ to perform calculations between shower development as a function of $X$, and atmospheric properties, which are functions of $z$.

\begin{figure}[t!]
\includegraphics[width=\linewidth]{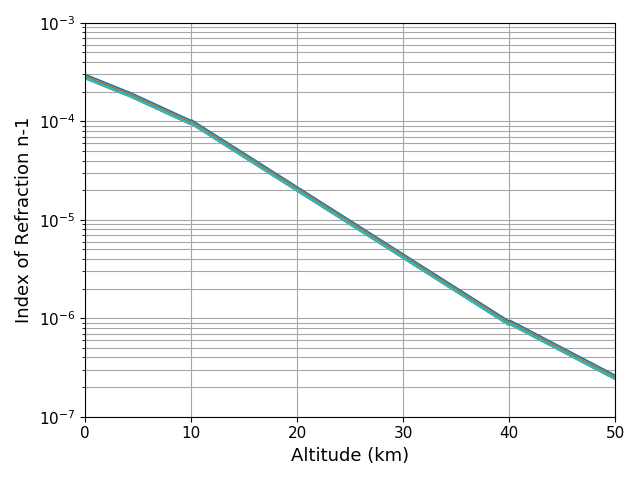}
\caption{Index of refraction (n-1) as a function of altitude for wavelengths 270~nm to 1000~nm. Because the difference is small in this wavelength range, dispersion effects are ignored.}
\label{RefractionIndex}
\end{figure}

Similarly, we use the index of refraction as a function of altitude and wavelength as given in CORSIKA \cite{Weast:1986uoy,Ciddor:1996aaa,Bernlohr:2008mpk}:

\begin{equation}
\begin{split}
 & n(z) =   1+0.283 10^{-3} \frac{\rho(z)}{\rho(0)}\\
 & n_{\lambda}(z) = 1+(n(z)-1) \left(0.967+0.033 \left(\frac{400}{\lambda (\mathrm{nm})}\right)^{2.5}\right)
\end{split}
\end{equation}
where $\rho(z)$ is the air density taken from the US standard atmosphere. The variation in the refraction index as a function of the wavelength is small within our relevant Cherenkov range, from here on defined to be in the interval from 270~nm to 1000~nm, as shown in figure \ref{RefractionIndex}. Therefore, in this work, we ignore the wavelength dependence of the index of refraction and use $n_{450~ \mathrm{nm}}(z)$ for all future calculations.

\begin{figure}[t!]
\includegraphics[width=\linewidth]{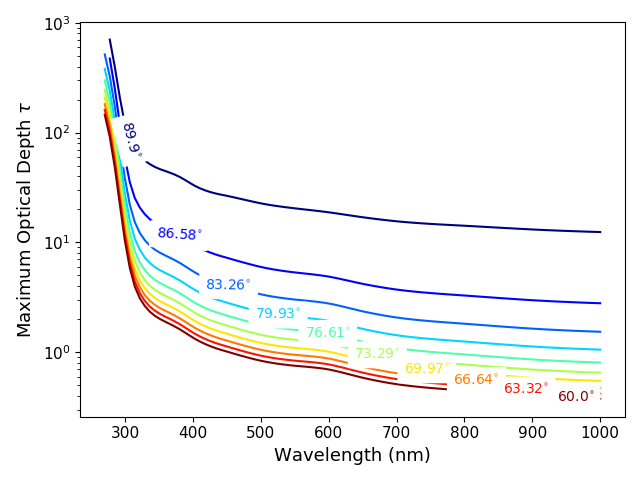}
\caption{Maximum optical depth as a function of wavelength for different trajectory angles (where $90^{\circ}$ corresponds to horizontal and $0^{\circ}$ corresponds to perfectly vertical).}
\label{OpticalThickness}
\end{figure}

Atmospheric extinction in the wavelength range 270~nm to 4~$\mu$m and altitude range 0~km to 50~km is given in \cite{Elterman:1968,Blitzstein:1970} where the attenuation coefficient (given by $\alpha(z,\lambda) = \sigma(\lambda) \cdot \rho(z) \cdot N_{avo}/m_{air}$) is provided for the Rayleigh, aerosol and ozone components both as analytical approximations and tabulated data in 1~km increments. Atmospheric transmission is calculated via:

\begin{equation}
\begin{split}
\tau(z,\lambda) &= \int_{z}^{z_{detector}} \alpha(z,\lambda) dl \\
\mathrm{Tr}(z,\lambda) &= e^{-\tau(z,\lambda)}\\
\end{split}
\label{Atmosphere_Extinction}
\end{equation}
where $dl$ is spaced in 10 ~m increments in the inherent spherical geometry of our coordinate system. We calculate optical depth using a 2-dimensional (wavelength and altitude) cubic interpolation on the tabulated data for the total scattering attenuation coefficient. Above 50~km altitude, we use the analytical cross section for Rayleigh scattering as an approximation for the total attenuation coefficient, given by: 
\begin{equation}
\sigma_{R}(\lambda) = \frac{24 \pi^{3}}{\lambda^{4}N^{2}} \left(\frac{n(\lambda)^{2}-1}{n(\lambda)^{2}+2}\right)^{2} F_{k}(\lambda)
\end{equation}
Where $N$ is the number density of air. In the case of air and the wavelength range relevant to us, $F_{k}(\lambda)$ is approximated as a constant 1.0608 \cite{Thalman:2014}. The maximum optical depth ($\tau(\lambda)|_{z = 0}^{z=100\mathrm{km}}$) as a function of wavelength for different trajectories is shown in figure \ref{OpticalThickness}. The ozone component of the atmosphere effectively attenuates light generated below 300~nm wavelengths.

\subsection{Optical Cherenkov Light Generation}
Charged particles in the shower generate Cherenkov photons while their energy is above the Cherenkov threshold, given by:

\begin{equation}
\begin{split}
&\beta>\frac{1}{n(z)} \text{ or, rearranging:}\\
&E>E_{thr} = \frac{m}{\sqrt{1-\frac{1}{n(z)^{2}}}}
\end{split}
\end{equation}
where $n(z)$ is the atmosphere's refraction index as function of the altitude (see figure \ref{RefractionIndex}) and $m$ the charge-particle mass.

The Cherenkov threshold energy is significantly larger for muons and hadrons than it is for $e^{\pm}$, as it scales with the particle mass. Additionally, the content of a cascade initiated by a proton, gamma, or $e^{\pm}$ is much more heavily composed of $e^{\pm}$ than other charged particles. For these reasons, we consider only the electron and positron component of the shower when calculating the Cherenkov emission, as the emission from other charged particle species is often subdominant \cite{Gaisser:2016uoy}. Whereas the particle content of the shower is computed with CORSIKA Monte Carlo simulations, we use shower parameterizations to model electron energy, angular and lateral distributions during the shower development. Although the parameterizations are given as function of shower age, which is independent of shower orientation, we note that these parameterizations are originally formulated only for downward going showers where Coulomb scattering is the dominant process forming the e$^\pm$ angular distributions and, as such, could show some deviations from the case of upward going showers. These deviations are expected to be small and in the present paper we will not touch this point.

The fraction of electrons and positrons above energy $E$ (expressed in MeV) in a given air shower as a function of shower age $s$ ($s = 3/[1+2(X_{\mathrm{max}}/X)]$) is given in \cite{Hillas:1982wz} as follows:

\begin{equation}
\begin{split}
T(E) &= \left(\frac{0.89E_{0}-1.2}{E_{0}+E}\right)^{s}(1+10^{-4}s E)^{-2}\\
E_{0} &= 
\begin{cases}
44-17(s-1.46)^{2},& \text{if } s\geq 0.4\\
    26,              & \text{otherwise}
\end{cases}
\end{split}
\end{equation}

\begin{figure}[t!]
\includegraphics[width=\linewidth]{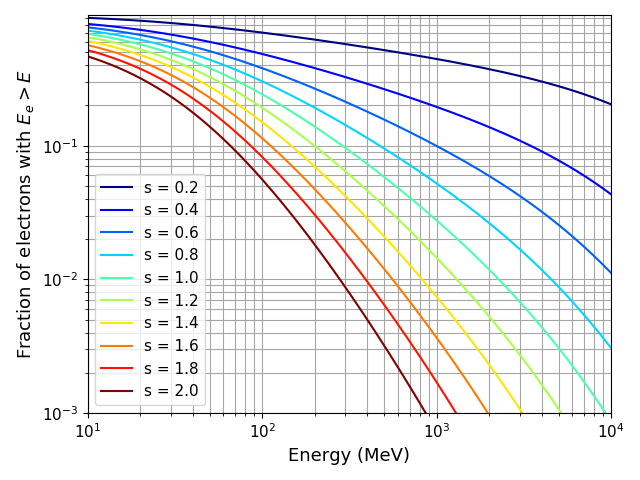}
\caption{Fraction of $e^{\pm}$ above energy $E$ for various shower ages, as parameterized in \cite{Hillas:1982wz}}
\label{AboveThreshold}
\end{figure}

$T(E)$ is plotted in figure \ref{AboveThreshold} for $s=0.2$ to $s=2.0$.

From our average longitudinal profile as computed with CORSIKA, we sample particle content in the shower in steps of 10~g/$\mathrm{cm}^{2}$ (or in 500 bins minimum, if the total grammage between the shower starting point and the detector is less than 5000~g/$\mathrm{cm}^{2}$) and calculate the corresponding altitudes and local Cherenkov threshold energy. For each bin in depth, we then divide the number of charged particles into 30 bins in energy from the local Cherenkov threshold up to $10^{7}$~MeV, spaced logarithmically. This upper limit was chosen so that even at extremely low shower ages (s=0.05), we ensure that over $99.99\%$ of the electron content is captured, regardless of the primary energy. The Cherenkov emission is calculated for 100 wavelength bins from 270~nm up to 1000~nm. From this, the number of Cherenkov photons generated in a grammage step $dX$ per unit wavelength $d\lambda$ and energy $d\mathrm{ln}E$ is given by:

\begin{equation}
E\frac{dN_{\gamma ch}}{dX d\lambda dE} =  \frac{N_{e^{\pm}}}{\rho(z)}\frac{ d T(E)}{d\mathrm{ln}E} \Big[2 \pi \alpha \frac{1}{\lambda^{2}} \Big(1-\frac{1}{(\beta n(z))^{2}} \Big) \Big]
\label{Cherenkov_Generation_eq}
\end{equation}
where $N_{e^{\pm}}$ is the number of $e^{\pm}$ in $d X$, $\alpha$ is the fine structure constant and $\rho(z)$ is the air density. To calculate the effects of atmospheric extinction, we multiply by the transmission function $\mathrm{Tr}(z,\lambda)$ defined in equation \ref{Atmosphere_Extinction}. Cherenkov photons are emitted in a ring with Cherenkov emission angle $\theta_{ch} = \mathrm{cos}^{-1}(1/\beta n)$ around the charged particle's trajectory. Since we use a fixed wavelength index of refraction $n_{450 \mathrm{nm}}$, $\theta_{ch}$ is a function only of electron energy and altitude. Thereby, we now integrate over wavelength to improve computational efficiency. Electrons in the shower propagate with an angular spectrum dependent on electron energy and minimally on shower age. For this, we use the Hillas parameterization \cite{Hillas:1982wz}, which gives the electron angular distribution in a universal form as follows:

\begin{equation}
\begin{split}
\frac{dn}{du} &= A e^{-(z-z_{0})/\lambda_{l}} \text{ where}\\
A &=
\begin{cases}
0.777,& \text{if } E < 350\mathrm{MeV}\\
    1.318, & \text{if }E>350\mathrm{MeV}
\end{cases}\\
z_{0} &=
\begin{cases}
0.59,& \text{if } E < 350\mathrm{MeV}\\
    0.37, & \text{if }E>350\mathrm{MeV}
\end{cases}\\
\lambda_{l} &=
\begin{cases}
0.478,& \text{if } z< z_{0} \text{ and } E<350 \mathrm{MeV}\\
    0.380, & \text{if }z>z_{0} \text{ and } E<350 \mathrm{MeV}\\
    0.413, & \text{if }z<z_{0} \text{ and } E>350 \mathrm{MeV}\\
    0.380, & \text{if }z>z_{0} \text{ and } E>350 \mathrm{MeV}
\end{cases}\\
\end{split}
\label{eq:AngularDist}
\end{equation}

where:

\begin{equation*}
\begin{split}
z &= \sqrt{u} \\
u &= \frac{w}{\langle w \rangle} \\
w &= 2(1-\mathrm{cos}\theta)(E/21~\mathrm{MeV})^{2})\\
\langle w \rangle &= 0.0054E(1+v)/(1+13v+8.3v^{2})\\
v &= E/E_{2}\\
E_{2} &= (1150+454 \mathrm{ln}(s)) (\mathrm{MeV})
\end{split}
\end{equation*}

The standard form is fairly complicated, but it is useful to note that $\frac{dn}{d\Omega} \propto \frac{dn}{du} \propto e(-\theta/\theta_{0}(E))$ where the scale angle $\theta_{0}$ is a function of the electron energy as discussed in \cite{Hillas:1982wz}. 

\begin{figure}[t!]
\includegraphics[width=\linewidth]{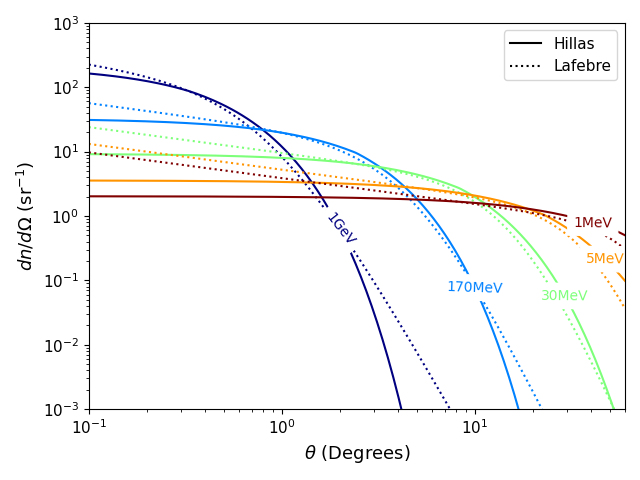}
\caption{Normalized angular distributions of electrons as a function of the angle off shower axis for electron energies from 1~MeV to 1~GeV for shower age $s=1.0$. As parameterized by Hillas \cite{Hillas:1982wz} and Lafebre \cite{Lafebre:2009en}. The former is used for the work presented in this paper.}
\label{AngDist}
\end{figure}

\begin{figure}[t!]
\includegraphics[width=\linewidth]{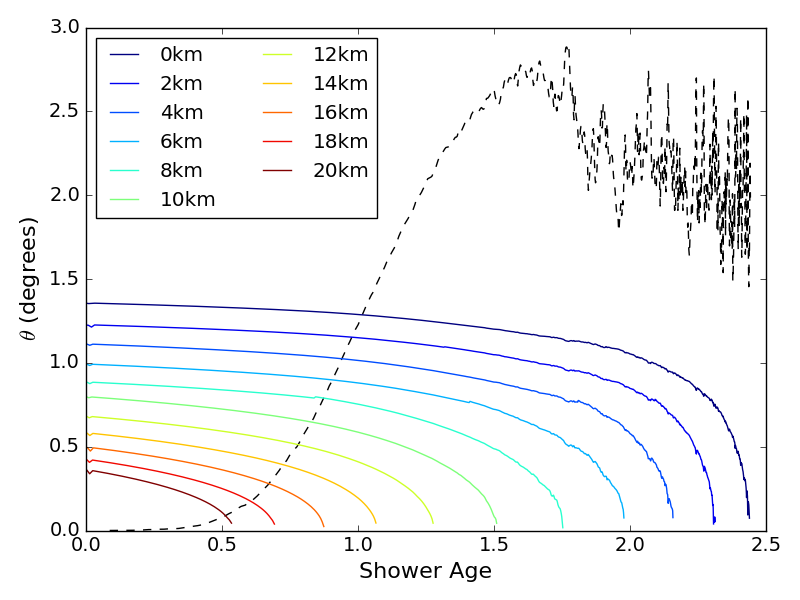}
\caption{Average electron scale angle (dashed) versus local Cherenkov angle as a function of shower age for a 100~PeV proton shower with $10^{\circ}$ Earth emergence angle initiated at the listed altitudes.}
\label{AngDistComp}
\end{figure}

Due to multiple scattering effects, higher energy electrons are distributed close to the shower propagation axis, whereas lower energy electrons have a larger angular spread. In figure \ref{AngDist}, we plot $\frac{dn}{d\Omega}$ as a function of the angle off shower axis $\theta$ for electron energies from 1~MeV up to 1~GeV at a shower age $s = 1.0$. It is helpful to qualitatively predict the behaviour of the Cherenkov signal, so we plot the average scale angle (calculated using the average electron energy at a given shower age) as a function of shower age for a given CORSIKA profile against the local Cherenkov angle for shower starting altitudes 0~km to 20~km in figure \ref{AngDistComp}.

For lower altitude shower development, the scale angle of the electron distribution and the local Cherenkov angle compete near shower maximum $s=1.0$. However, for showers initiated at higher altitudes, the electron scale angle begins to dominate the shower behaviour. Thus, for EAS which develop deep (higher altitude for upward-going EASs) in the atmosphere, we expect them to lose features from the Cherenkov ring and to more closely follow the exponential behaviour of the underlying electron angular distribution. In general, for higher altitude shower development, both angular scales are small, resulting in a more focused Cherenkov light distribution overall. We also include the effects of the lateral distribution of electrons in a shower as discussed in \cite{Lafebre:2009en} and given by:

\begin{equation}
\begin{split}
\frac{dn}{d\mathrm{ln}E d \mathrm{ln} x} &= C_{2}'x^{\zeta_{0}'}(x_{1}'+x)^{\zeta_{1}'}\\
x_{1}' &= 0.859-0.0461 \mathrm{ln}^{2}E + 0.00428 \mathrm{ln}^{3}E\\
\zeta_{t} &= 0.0263t\\
\zeta_{0}' &= \zeta_{t}+1.34+0.160 \mathrm{ln}E - 0.0404 \mathrm{ln}^{2}E \\
&+ 0.00276 \mathrm{ln}^{3}E\\
\zeta_{1}' &= \zeta_{t} - 4.33
\end{split}
\end{equation}
where $C_{2}'$ is the normalization constraint, $t = (X-X_{\mathrm{max}})/X_{0}$, $X_{0}$ being $36.7 \mathrm{g}/\mathrm{cm}^{2}$, the air radiation length of electrons, and $x = r/r_{m}$ with $r_{m}$ the Moliere radius given by $9.6 ~\mathrm{g} \, \mathrm{cm}^{-2}/\rho(z)$ \cite{Lafebre:2009en}. We plot both the lateral distribution as given in \cite{Lafebre:2009en} for the electron energy range (1~MeV, 1~GeV) at shower maximum and the Moliere radius as a function of altitude, to quantify the scale of this process.

\begin{figure}[t!]
\includegraphics[width=\linewidth]{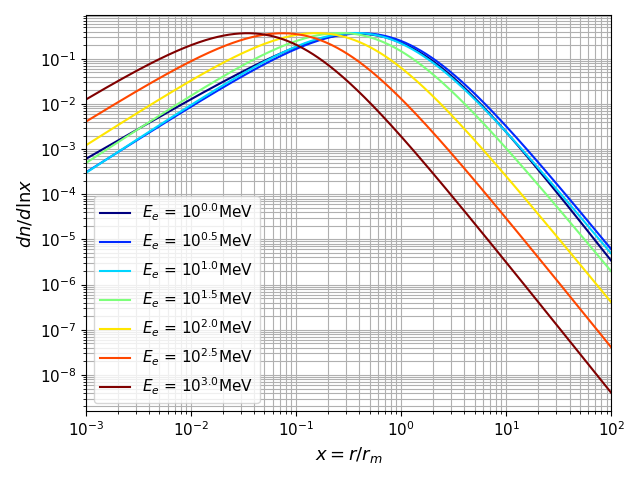}
\caption{Normalized electron lateral distributions for electron energies 1~MeV to 1~GeV at shower maximum.}
\label{LafebreDist}
\end{figure}

\begin{figure}[t!]
\includegraphics[width=\linewidth]{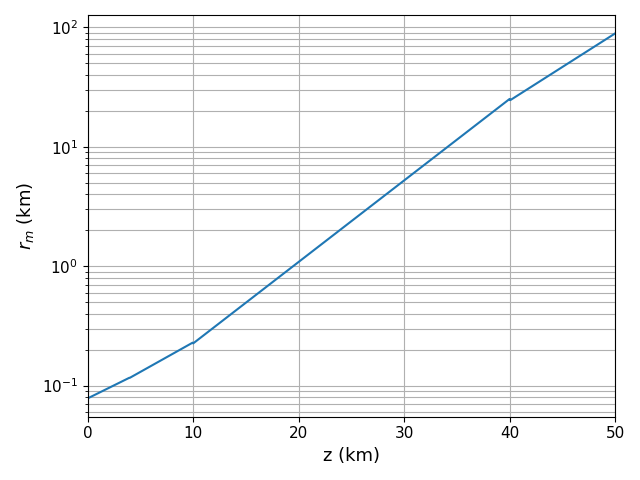}
\caption{Moliere radius as a function of altitude for the standard US atmosphere.}
\label{MoliereRadius}
\end{figure}

The characteristic length scale for the Cherenkov signal transverse width from an upward-moving EAS air shower initiated at sea level is ${\rm h} \times \mathrm{tan}(\theta_{ch})$, where $h$ is the altitude of the detector and $\theta_{ch}$ the Cherenkov angle introduced before, which is $1.4^{\circ}$ at sea level. In the cases we are interested in: POEMMA and EUSO-SPB2, the altitudes are respectively $h=535$~km and $h=33$~km. Therefore, the characteristic Cherenkov length scale will be $\mathcal{O}$$(10~\mathrm{km})$ in the case of POEMMA and $\mathcal{O}$$(1~\mathrm{km})$ in the case of EUSO-SPB2.

The lateral distribution of electrons becomes a relevant process when the lateral scale $r$ competes with the Cherenkov length scale, which more readily occurs for high altitudes and low electron energies. For a space based instrument, this process is largely negligible, as the Moliere radius is larger than 10~km only above 30~km altitude, where Cherenkov generation is minimal. For a balloon based instrument, however, the lateral spreading of electrons is a prominent effect for two reasons: the Moliere radius approaches the Cherenkov scale at lower altitudes where shower development is ongoing, and the detector can be located within the developing shower, for EAS initiated at higher altitudes, e.g. EeV-scale \taon decay or muon interaction, where the lateral distribution becomes the dominant effect on the spatial distribution of Cherenkov photons.

To consider the effects of the angular and lateral electron distributions as described above, we sample 200 bins of $u$ (see equation \ref{eq:AngularDist}) from $10^{-4}$ to $1$ using logarithmic sampling, and calculate $dN_{\gamma}/d\mathrm{ln}E dX du$ and the corresponding zenith angle off shower axis according to equation \ref{eq:AngularDist}. Using the zenith angle, the local distance to detector plane, and the Cherenkov angle, the Cherenkov ring is projected onto an ellipse in 500 angular bins. To save on computational time and complexity, we do not model the entire lateral distribution to describe the behavior of the shower. Instead, the coordinates of each point on the ellipse are shifted by a randomly sampled angle $\phi$, and displacement $r$, sampled from the lateral distribution of a given electron energy, shower age, and the local Moliere radius. To account for azimuthal symmetry of the shower, the distance from shower axis is calculated for each bin coordinate and randomly rotated about the shower axis. 

\begin{figure}[t!]
\includegraphics[width=0.87\linewidth]{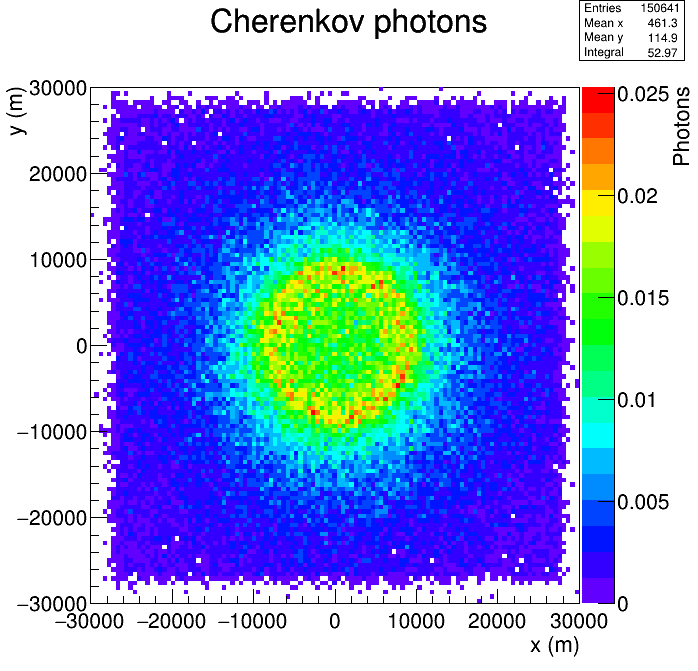}
\includegraphics[width=\linewidth]{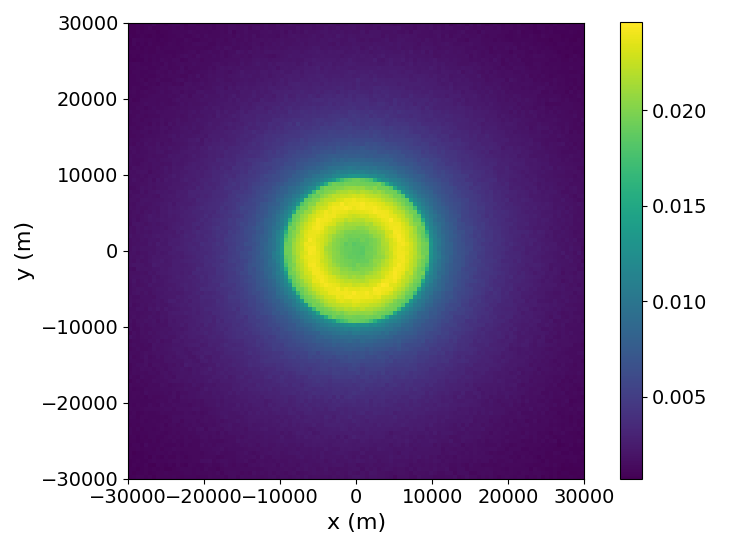}
\caption{Cherenkov spatial distributions for an upward ($90^{\circ}$ Earth emergence angle) 10~TeV proton shower as observed at 400~km altitude, with $\lambda$ range 300~nm to 450~nm, simulated with CORSIKA [upper panel] and this work [lower panel]. Color bar represents $N_{\gamma_{ch}}/m^{2}$ and is scaled by $\frac{1}{10}$ to decrease computation time in CORSIKA for projecting photons from top of the atmosphere to 400~km. CORSIKA run provided by F. Bisconti.}
\label{fig:CORSIKA_Comp}
\end{figure}

The photons are spatially sampled on the detector plane and integrated in $X$, $\mathrm{ln}E$ and $u$ to yield the spatial distribution of the Cherenkov signal from the shower. A modified version of CORSIKA-75600 was developed to produce the optical Cherenkov emission from an upward going air shower in a flat atmosphere \cite{ThanksDieter}. For most of the geometries we simulate, we cannot reliably approximate the atmosphere as flat, but for a perfectly vertical air shower, our results agree quite well with the CORSIKA results, as shown in figure \ref{fig:CORSIKA_Comp}.

\begin{figure}[t!]
\includegraphics[width=\linewidth]{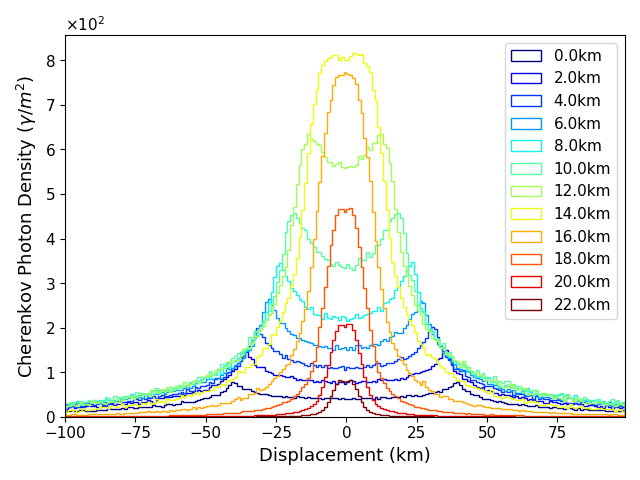}
\includegraphics[width=\linewidth]{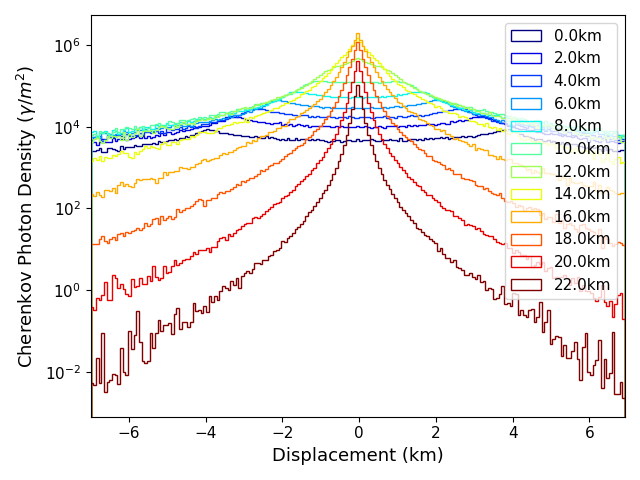}
\caption{Cherenkov spatial distributions for an upward 100~PeV proton air shower at $10^{\circ}$ Earth emergence angle for starting altitudes from 0~km to 22~km, as observed by a space based instrument at 525~km [upper panel] and a balloon based instrument at 33~km [lower panel]. The spatial distribution for the balloon based instrument is plotted in logarithmic scale to better demonstrate the differences between the curves.}
\label{fig:Example_Dists}
\end{figure}

In figure \ref{fig:Example_Dists}, we plot the Cherenkov spatial distributions for the showers plotted in figure \ref{AngDistComp} as observed by POEMMA ($h=525$~km) and EUSO-SPB2 ($h=33$~km) to qualitatively describe features of these signals. As expected, the Cherenkov light pool for POEMMA has a lateral scale $\mathcal{O}$(10~km) and that of EUSO-SPB2 has $\mathcal{O}$(1~km). The difference in geometry alone leads to an $\mathcal{O}$(100) difference in central photon intensity between the two, which can best be observed by comparing the showers with 0~km starting altitude. With increasing shower starting altitude, the spatial extent of the showers shrinks, and the characteristic Cherenkov "horns" diminish, due to the decreasing Cherenkov angle and electron scale angle as described above. 

With increasing starting altitude, showers experience forward beaming and decreased atmospheric attenuation, mainly from minimizing the effects of aerosol attenuation, leading to great increases in photon intensity. Above 15~km starting altitude, photon intensity begins to decrease due to the suppression of Cherenkov photon generation from the increased Cherenkov threshold. We also note the very large disparity in Cherenkov intensity between low and high altitude showers observed from balloon altitudes due to primarily near-field effects which are absent from the Cherenkov spatial profiles observed at low-Earth orbit and higher altitudes.

\begin{figure}[t!]
\includegraphics[width=\linewidth]{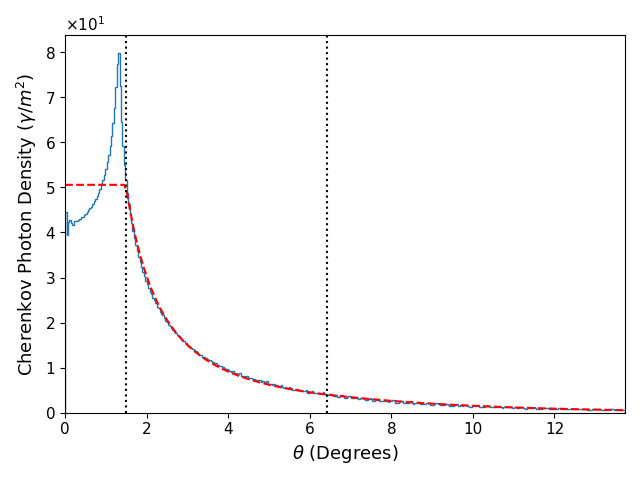}
\includegraphics[width=\linewidth]{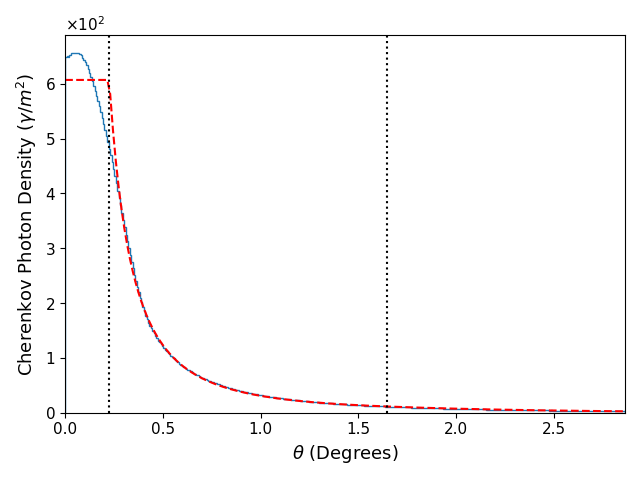}
\caption{[Upper panel] Cherenkov light distribution for 100~PeV proton shower with $10^{\circ}$ Earth emergence angle initiated at 0~km altitude as seen by POEMMA, with profile fits described in the text. Vertical lines show the transition from constant to power law to exponential behaviors. [Lower panel] Shower initiated at 17~km altitude.}
\label{ProfileFits}
\end{figure}

\subsection{1D Shower Modeling}
The computation of the three dimensional Cherenkov emission produced by a single shower takes a significant amount of computational resources. It is thus useful to model the three dimensional output of the Cherenkov spatial light profile analytically using a one dimensional profile of the Cherenkov photon density (in photons per $\mathrm{m}^{2}$) as a function of $\theta$, the angle off shower axis as measured from the ground, as follows: 


\begin{equation}
\rho_{ch} = 
\begin{cases}
\rho_{0} & \theta \leq \theta_{ch}\\
\rho_{0}\Big(\frac{\theta}{\theta_{ch}}\Big)^{-\beta} & \theta_{ch} \leq \theta \leq \theta_{1}  \\
\rho_{1}e^{-(\theta-\theta_{1})/\theta_{2}} & \theta \geq \theta_{1}
\end{cases}
\label{1dprof}
\end{equation}


Where the free parameters are $\rho_{0}$, the central Cherenkov intensity in photons/$\mathrm{m}^{2}$, $\theta_{ch}$, the angle within which the Cherenkov spatial distribution is approximated as flat, which corresponds well to the Cherenkov angle at $X_{\mathrm{max}}$ projected from ground, $\beta$, the power law index describing the behavior of the tails close to the shower axis, $\theta_{1}$ the transition angle from power law to exponential behavior, and $\theta_{2}$, the exponential scale of the tails far off shower axis. $\rho_{1}$ is calculated such that the function remains continuous between the last 2 regions, and is not a free parameter. For a given shower geometry (trajectory angle and shower starting altitude), we perform a least squares fit to the Cherenkov spatial distribution with the 5 parameter fit described above. The results of the fits are shown in figure \ref{ProfileFits} for a 100~PeV proton shower with Earth emergence angle $10^{\circ}$ and starting altitudes 0~km and 17~km. The flat-top fit to the distribution provides an effective average close to the shower axis, while accurately modeling the tails of the distribution out to very large angles off of shower axis.

\begin{figure}[t!]
\includegraphics[width=\linewidth]{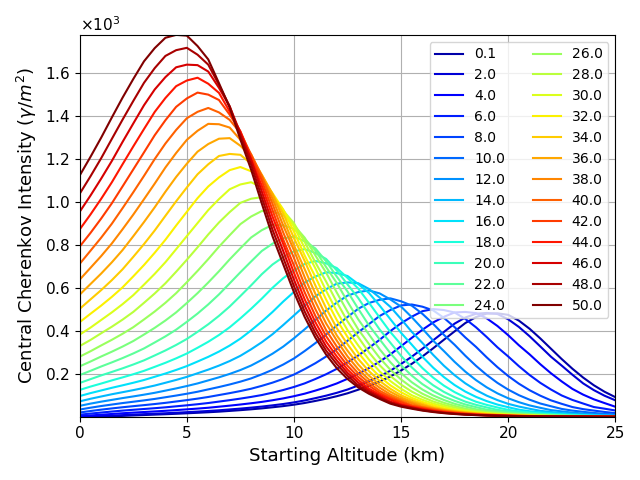}
\includegraphics[width=\linewidth]{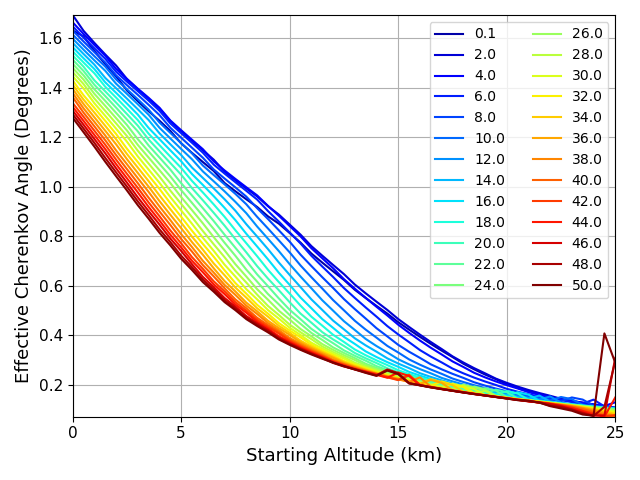}
\includegraphics[width=\linewidth]{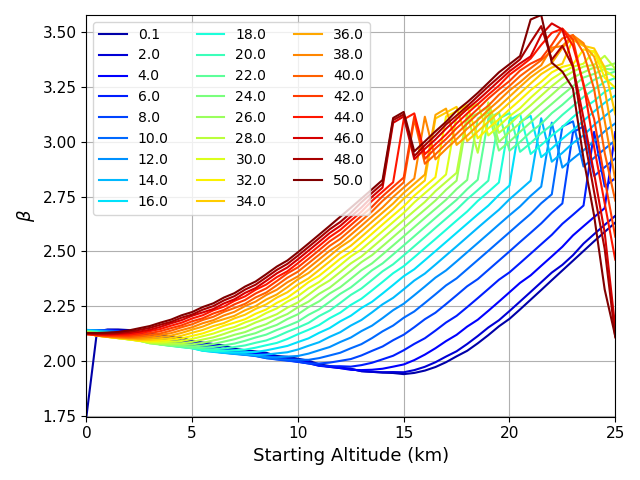}
\caption{Parameter fits to the Cherenkov spatial distribution from a 100~PeV upward proton shower initiated at starting altitudes 0~km to 25~km and Earth emergence angles $0^{\circ}$ to $50^{\circ}$ as observed by a space based instrument at 525~km altitude. Plots are central intensity $\rho_{0}$ [upper panel], central width $\theta_{ch}$ [middle panel], and power law scale $\beta$ [lower panel] using the 5 parameter fit model described in the text.}
\label{fig:POEMMA_Params}
\end{figure}

\begin{figure}[t!]
\includegraphics[width=\linewidth]{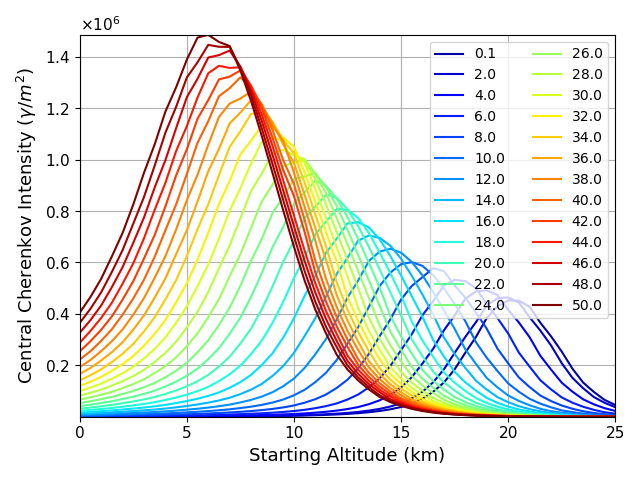}
\includegraphics[width=\linewidth]{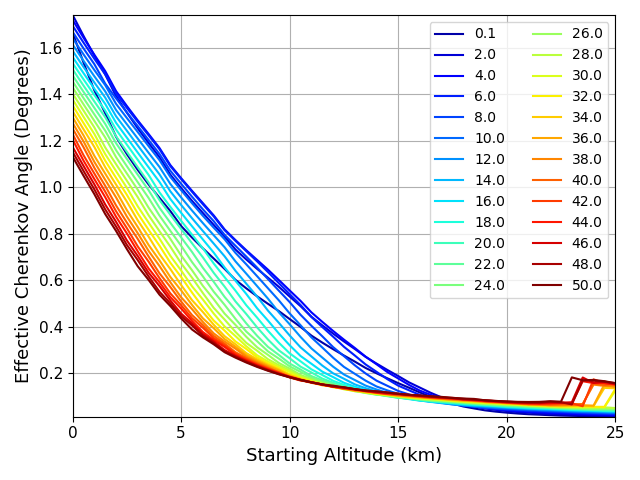}
\includegraphics[width=\linewidth]{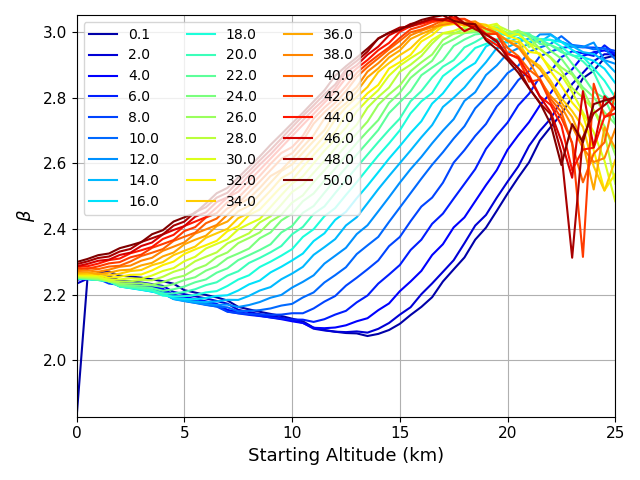}
\caption{Parameter fits to the Cherenkov spatial distribution from a 100~PeV upward proton shower initiated at starting altitudes 0~km to 25~km and Earth emergence angles $0^{\circ}$ to $50^{\circ}$ as observed by a balloon based instrument at 33~km altitude. Plots are central intensity $\rho_{0}$ [upper panel], central width $\theta_{ch}$ [middle panel], and power law scale $\beta$ [lower panel] using the 5 parameter fit model described in the text.}
\label{fig:SPB_Params}
\end{figure}


Previous attempts to fit the Cherenkov spatial distribution used solely power law or exponential models to describe the off axis behavior, but this results in large overestimates and underestimates, respectively. This combined fit presented in figure \ref{ProfileFits} describes the Cherenkov photon distribution quite well, even out to very large angles off shower axis, which becomes important for showers with very large primary energies.

To generate a large parameter space on which to sample, we simulate showers with starting altitudes from 0~km to 25~km in 0.5~km increments, and Earth emergence angles from $0.1^{\circ}$ to $50^{\circ}$ in $2^{\circ}$ increments. In figures \ref{fig:POEMMA_Params} and \ref{fig:SPB_Params}, we plot the three dominant parameters of the fit (the central intensity $\rho_{0}$, the effective Cherenkov width $\theta_{ch}$, and the log scale $\beta$, which describe the shape of the distribution close to the axis) to the Cherenkov spatial distribution for a space based instrument and a balloon based instrument. The relative intensity of the Cherenkov signal for a 100~PeV upward going EAS is roughly 3 orders of magnitude larger for a balloon based instrument than a space based instrument. The difference in intensities can be reasonably approximated as the ratio of $\big(\frac{L_{525}}{L_{33}} \big)^{2}$ (where $L$ here is the path length from $X_{max}$ to observation), but breaks down when $L$ is of the same magnitude as $r_{m}$ the Moliere radius, which occurs for balloon based instruments, making them more sensitive to near-field effects. 

The effective Cherenkov angle $\theta_{ch}$ and the scale $\beta$ of the Cherenkov spatial profile decrease and increase, respectively, with increasing starting altitude, resulting in a very narrow, steeply falling distribution. However, for very high starting altitudes ($z \geq \,$20~km), we begin to see the opposite behavior, that is, a slight rebound in $\theta_{ch}$, and a decrease in $\beta$. These are the effects of the electron lateral distribution, which result in a more spatially spread signal. As expected, the effects of the lateral distribution become important earlier for balloon-based altitudes, as an instrument observing at these altitudes can often be inside the active shower development.


\begin{figure}[t!]
\includegraphics[width=\linewidth]{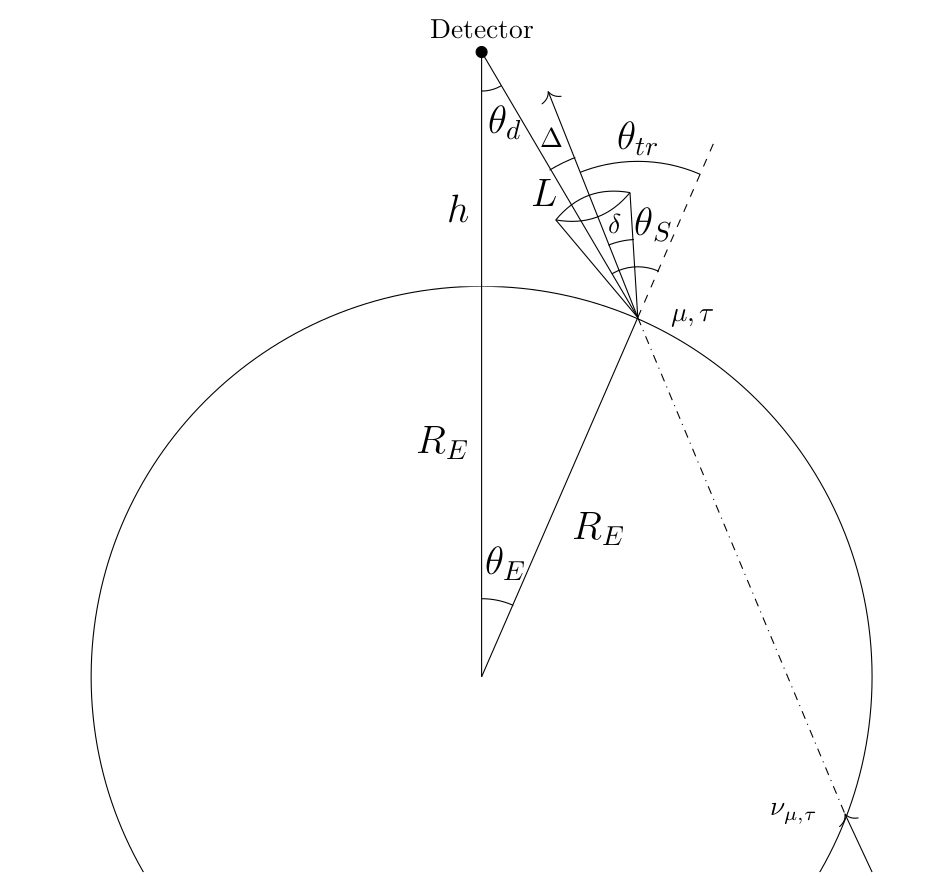}
\caption{Geometry of a neutrino induced upward going air shower as observed by a detector at altitude h (see text for details).}
\label{Geometry}
\end{figure}

\section{Aperture and Sensitivity}
\label{rates}

\subsection{Geometric Aperture}
The geometry of an upward EAS produced by the decay ($\tau$-lepton) or interaction (muon) at the Earth surface is sketched in figure \ref{Geometry}. The distances $h$, $L$, $R_E$ are the altitude of the detector, the path length from the emission point to the detector, and the Earth's radius. The three angles $\theta_{S}$, $\theta_{d}$ and $\theta_{E}$, related as $\theta_d=\theta_{S}-\theta_E$, are respectively: the angle of the detector's line of sight (respect to the local zenith), the detector's viewing angle and the Earth viewing angle. The angle $\delta$ is the critical angle within which the Cherenkov emission from the EAS can trigger the instrument and the angle $\Delta$ is the angular displacement between the shower direction and the detector's line of sight. The Earth emergence angle is given by $\theta_{sh}=\frac{\pi}{2}-\theta_{tr}$, being $\theta_{tr}$ the angle between the local zenith and the shower direction (see figure \ref{Geometry}).

To calculate the geometric aperture of orbital or sub-orbital detectors, we refer to the general discussion presented in \cite{Motloch:2013kva}. Using the conventions highlighted in figure \ref{Geometry}, the geometric aperture to detect upward going showers is given by:

\begin{equation}
\begin{split}
& \braket{A\Omega} (E_\nu) = \int_{\Delta S} \int_{\Delta \Omega_{det}} P_{obs} \hat{r} \cdot \hat{n} dS d\Omega_{det} = \\
& = R_E^2 \int_{\Delta \Omega_E} \int_{\Delta \Omega_{det}} P_{obs}(E_{\nu},\theta_{tr}) \hat{r} \cdot \hat{n} d \Omega_{E} d \Omega_{det}
\label{aperture1}
\end{split}
\end{equation}
where $P_{obs}(E_{\nu},\theta_{tr})$ is the combined neutrino emergence probability, lepton decay probability and decay channel probability, $\hat{n}$ corresponds to the Earth normal and $\hat{r}$ to the shower trajectory, $\Delta S = R_E^2 \Delta\Omega_E$ is the area of the spherical zone on ground visible to the detector and $\Delta \Omega_{det}$ is the solid angle of detection of the Cherenkov emission. Using the coordinate system in which the detector line of sight ($L$ in figure \ref{Geometry}) points in the $\hat{z}$ direction and the Earth normal $\hat{n}$ lies in the $\hat{y}\hat{z}$ plane, we can rewrite the scalar product $\hat{r}\cdot \hat{n}=\mathrm{cos}\theta_{S} \mathrm{cos}\Delta+\mathrm{sin}\theta_{S}\mathrm{sin}\Delta \mathrm{sin}\phi_{p}$ and $d\Omega_{det}=d(\mathrm{cos} \Delta)d\phi_p$, being $\phi_p$ the azimuth angle of the EAS's trajectory around $\hat{z}$. Integrating over the azimuth angle $\phi_p$ and retaining only the (dominant) emission probability along the detector's line of sight $P_{obs}(E_\nu,\theta_{tr}) = P_{obs}(E_\nu,\theta_s-\Delta)\simeq P_{obs}(E_\nu,\theta_s)$, one has:

\begin{equation}
\begin{split}
& \braket{A\Omega}  (E_{\nu})  = 2 \pi R_{E}^{2} \times \\
& \times  \int_{\Delta\Omega_E} \int_{\mathrm{cos}\delta}^1 P_{obs}(E_\nu,\theta_{tr}) \mathrm{cos} \theta_{S} \mathrm{cos}\Delta d(\mathrm{cos}\Delta) d\Omega_E = \\
& = \pi R_{E}^{2}  \int_{\Delta\Omega_E} P_{obs}(E_{\nu},\theta_{S}) \mathrm{cos} \theta_{S} [1-\mathrm{cos}^2\delta]d\Omega_E
\end{split}
\label{aperture} 
\end{equation}
where $\delta=\delta(\theta_E)$ is the maximum angle off shower axis at which the Cherenkov spatial distribution is still observable (see equation \ref{delta}) and $\mathrm{cos} \theta_{S}$ follows from the geometry (see figure \ref{Geometry}) as:

$$ 
\mathrm{cos} \theta_{S} = \frac{(R_E+h)\mathrm{cos} \theta_{E} - R_E}{\sqrt{R_E^2 \mathrm{sin}^2 \theta_{E}  + [R_E+h-R_E\mathrm{cos} \theta_{E}]^2 }}~.
$$

The integration over $d\Omega_E=d(\mathrm{cos}\theta_E)d\phi_E $ should take into account the observational limits of the detector, in both the zenith $\theta_E$ and azimuth $\phi_E$ angles. The integration over $\phi_E$ is straightforward and it gives: 

\begin{equation}
\begin{split}
& \braket{A\Omega} (E_{\nu}) =\\
&= \pi R_{E}^{2} \Delta\phi_E \int P_{obs}[E_\nu,\theta_s(\theta_E)] \times \\
& \times \mathrm{cos}[\theta_{S}(\theta_E)] \mathrm{sin}^{2}[ \delta(\theta_{E})] \mathrm{sin}\theta_{E} d\theta_{E}\end{split}
\label{aperture2}
\end{equation}
where $\Delta\phi_E$ is the range in the azimuth angle visible by the detector. In the following we will consider the ideal case $\Delta\phi_E=2\pi$ and, taking into account the actual experimental designs, the results are later scaled by the azimuth ranges of $\Delta\phi_E=30^{\circ}$ and $\Delta\phi_E=12.8^{\circ}$ for POEMMA and EUSO-SPB2 respectively. Using the viewing angle ranges for $\Delta \theta_d$ of $7^{\circ}$ below the Earth's limb for POEMMA and $6.4^{\circ}$ for EUSO-SPB2 \cite{Reno:2019jtr,Anchordoqui:2019omw,Olinto:2019mjh,Wiencke:2019vke,Adams:2017fjh}, the limits on $\theta_E$ in equation \ref{aperture2} are calculated. 

\subsection{Monte Carlo Methodology}
To calculate $\delta(\theta_{E})$ for different primary neutrino energies, we developed a simple Monte Carlo code which estimates the average optical Cherenkov signal for a given neutrino flux. For Earth emergence angles in the range $(0^{\circ},50^{\circ})$ in $2^{\circ}$ increments and primary neutrino energies between 1~PeV and 100~EeV in 0.1 decade increments, we simulate $10^{5}$ 1D Cherenkov profiles for showers induced by 1) hadrons and electrons from \taon decay, 2) secondary muons from \taon decay, 3) primary muons from muon neutrinos interacting in the Earth. These signals can then be used to calculate the sensitivity of the instruments to an isotropic diffuse neutrino flux, and thereby the expected neutrino event rate. We perform this process for the proposed space-based POEMMA mission ($h=525$ km) and for the balloon-borne Cherenkov telescope in EUSO-SPB2 ($h=33$ km) currently under construction.

\begin{figure}[t!]
	\includegraphics[width=\linewidth]{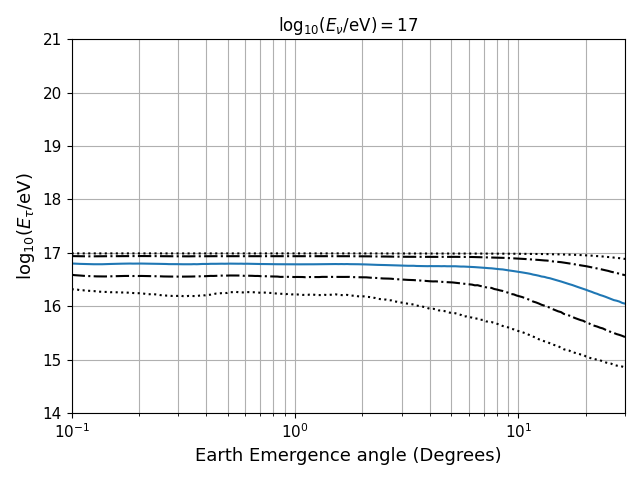}
	\caption{Emerging \taon energy distribution for a $10^{17}$eV parent neutrino as a function of Earth emergence angle calculated using 3-dimensional kernel density estimation. The center blue line represents the mean \taon energy and the dash-dotted and dotted lines represent the $1\sigma$ and $2\sigma$ deviations, respectively.}
	\label{fig:KDE_Example}
\end{figure}

Kernel density estimation is used to generate a 3-dimensional probability distribution function from 100 discrete emerging lepton energies over the fractional energy range ($10^{-6},1$) for the neutrino energies $E_\nu$ and Earth emergence angles $\theta_{sh}$ simulated in NuTauSim \cite{Alvarez-Muniz:2017mpk}. Moreover, for a given couple $(E_\nu,\theta_{sh})$, a 1-dimensional probability distribution function is generated and sampled to determine emerging lepton energies to start the propagation. An example output of this 3-dimensional Kernel density estimation for a 100~PeV $\tau$ neutrino is shown in figure \ref{fig:KDE_Example}.

If the Earth-emergent particle is a $\tau$-lepton, the decay length in the atmosphere is selected randomly from an exponential distribution with mean $4.9 \times (E_{\tau}/10^{17}$~eV) km, and the starting altitude $z_{st}$ is found using the spherical geometry for $\theta_{E}$. 

To account for the selection of the decay channels and the possibility of decay outside the interaction range, the event is given a probability weight $\Gamma^\tau_{e^{-},had}(1-e^{-L/L_{d}})$ where $L$ is the path length to a maximum predetermined altitude (which we choose to set to 25~km as negligible EAS development occurs above), $\Gamma^\tau_{e^{-},had}=0.812$ is the \taon decay branching ratio into electrons and hadrons and $L_{d}$ is the average decay length of the \taon with sampled energy $E_{\tau}$. 

The fractional shower energy $\eta$ is randomly sampled from the PYTHIA distributions of a tau decay shown in figure \ref{Energy_Dists} and the effective shower energy $E_{sh} = \eta E_{\tau}$ is calculated, where $\eta$ is sampled from all decay channels which result in either hadrons or electrons being emitted. Using $\theta_{E}$ and $z_{st}$, we interpolate the Cherenkov profile parameters from figure \ref{fig:POEMMA_Params} for POEMMA or \ref{fig:SPB_Params} for EUSO-SPB2, and scale the intensity to $E_{sh}/10^{17}\mathrm{eV}$.

If the EAS generating particle is a secondary muon from \taon decay, the muon energy $E_{\mu}$ is sampled from the muon decay channel from PYTHIA \cite{Sjostrand:2014zea}, with a probability weight (branching ratio) $\Gamma^\tau_{\mu}=0.173$.  If the EAS is by an Earth-emergent muon from a muon neutrino interaction, the muon sampling is performed as discussed above, using the results of the muon modification to NuTauSim \citep{Romero-Wolf:2018zxt}. The total atmospheric slant depth $X_{\mathrm{tot}}^{\mu}$ from the muon origin point to 25~km altitude is calculated, and the corresponding lower bound on the fractional energy $\eta$ is found such that $X_{\mathrm{tot}}^{\mu}=5X_{\mathrm{int}}^{\mu}$ ($99\%$ of particles interact within 5 interaction lengths), where $X_{\mathrm{int}}^{\mu}$ is given by equation \ref{eq:Xint}. The muon (either from \taon decay or $\nu_\mu$ CC interaction) is propagated through the atmosphere in steps of $dX$ (sampled from an exponential distribution with mean $X_{\mathrm{int}}$) until exiting and, at each step, the fractional energy $\eta$ is sampled from the differential muon cross sections. We use the differential and integrated cross section of a 100~PeV muon for all of our calculations, as the muon cross section is weakly dependent on energy for $E_{\mu}>1 \mathrm{TeV}$. The deposited energy which begins an air shower $E_{sh} = \eta E_{\mu}$ is logged and subtracted from the propagating muon energy. For every interaction, the instantaneous altitude is calculated and the Cherenkov profile parameters are interpolated and scaled by $E_{sh}/10^{17}\mathrm{eV}$. 

From the interpolated Cherenkov spatial distribution, and given a photon detection threshold $\rho_{thr}$ (the photon density required to disentangle the Cherenkov signal from the background--which is discussed further in the following section), we then calculate the angle off axis $\delta$ to which the signal would still be observable for our instrument. $\delta$ is given by inverting the 1D Cherenkov model in equation \ref{1dprof}:

\begin{equation}
\begin{split}
\delta &=
\begin{cases}
0 & \rho_{0} \leq \rho_{thr}\\
\theta_{ch} \Big( \frac{\rho_{thr}}{\rho_{0}}\Big)^{-1/\beta} & \rho_{1} \leq \rho_{thr} \leq \rho_{0}\\
\theta_{2}\mathrm{ln}\Big( \frac{\rho_{1}}{\rho_{thr}}\Big)+\theta_{1} & \rho_{thr} \leq \rho_{1}
\end{cases}
\end{split}
\label{delta}
\end{equation}

where the scales $\theta_{ch}$, $\theta_{1}$, $\theta_{2}$, and $\beta$ are defined above (see equation \ref{1dprof}), and $\rho_{0}$ is the central density of the one dimensional profile. For each ($E_{\nu},\theta_{sh}$), we calculate the average $\delta(\theta_{E})$ across all $10^{5}$ thrown showers. 

Note that for a muon initiated EAS with multiple interactions along the muon trajectory, we track only the spatial profile of the interaction which provides the largest angle to which the instrument is sensitive. That is, we do not calculate the superposition of spatial distributions from multiple interactions of the muon, but only that from the interaction which provides the largest signal at the detector. This was found to have a minimal ($<5\%$) effect on the maximum observable angle from shower axis.

\subsection{Event Rate Estimations}
\label{events}
The photon detection threshold of the two experiments we are concerned with is given by $\rho_{thr}=N^{min}_{PE}/(\epsilon_Q A)$, where $\epsilon_Q$ is the quantum efficiency for background photons detection, $A$ is the effective photon collecting area and $N^{min}_{PE}$ is the minimum number of photo-electrons corresponding to a fixed maximum rate of "false events" triggered by the background (here fixed below $10^{-2}$ per year). 

As discussed in \cite{Reno:2019jtr}, integrating the signal into the typical duration of a Cherenkov burst (20 ns), the reference value $N_{PE}^{min}=10$ (40 for EUSO-SPB2) can be used as it follows from the technical design of POEMMA and the dark-sky air glow background model in the 300-1000 nm range. This model, based on the VLTL/UVES measurements \cite{2003A&A...407.1157H,2006JGRA..11112307C} with the van Rhijn enhancement \cite{VanRhijn:1921,1955ApJ...122..530R,1965P&SS...13..855B}, implies a quantum efficiency for the background detection with $\epsilon_Q=0.2$, assuming the same performances of the Cherenkov signal detection \cite{Reno:2019jtr,Anchordoqui:2019omw}. The photon collecting area of POEMMA and EUSO-SPB2 are respectively $A=2.5$ m$^2$  and $A=1$ m$^2$ and the threshold photon density $\rho_{thr}$ will be respectively 20 photons/m$^2$ and 200 photons/m$^2$.

\begin{figure}[t!]
\includegraphics[width=\linewidth]{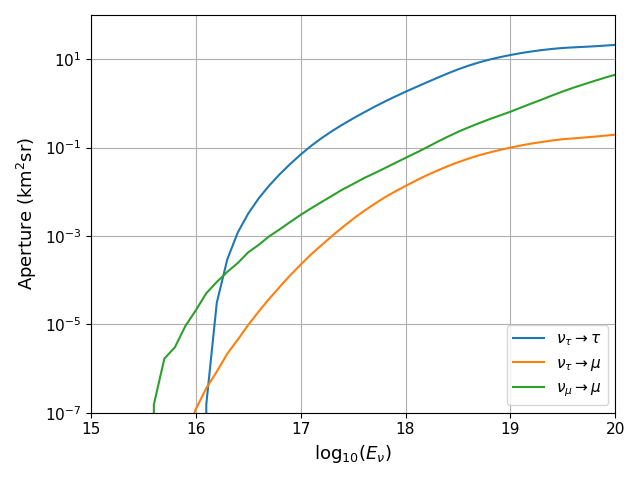}
\includegraphics[width=\linewidth]{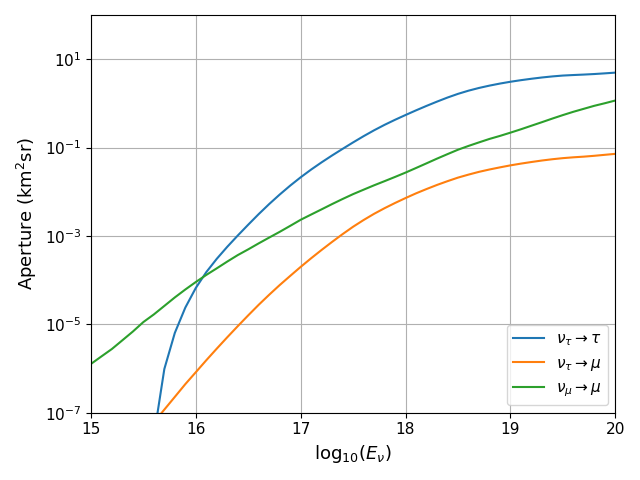}
\caption{Geometric aperture as a function of primary neutrino energy for the primary tau neutrino, primary and secondary muon neutrino detection channels for POEMMA-$360^{\circ}$ [upper panel] and EUSO-SPB2-$360^{\circ}$ [lower panel].}
\label{fig:apertures}
\end{figure}

To determine the geometric aperture $\langle A\Omega \rangle (E_\nu)$ (where the nominal $20\%$ duty cycle of each instrument is not yet applied) to an isotropic neutrino flux, we perform the numerical integration of equation \ref{aperture2} over the spatial region covered by the fields of view of POEMMA and EUSO-SPB2. The angular parameter $\delta$ is sampled from the simulated Cherenkov showers, as before, using the detailed fit to the 1D Cherenkov spatial distribution. The resulting geometric aperture of POEMMA-like and EUSO-SPB2-like instruments (unscaled by the proper azimuth ranges) for the 3 detection channels as a function of primary neutrino energy is shown in figure \ref{fig:apertures}.

\begin{figure}[t!]
\includegraphics[width=\linewidth]{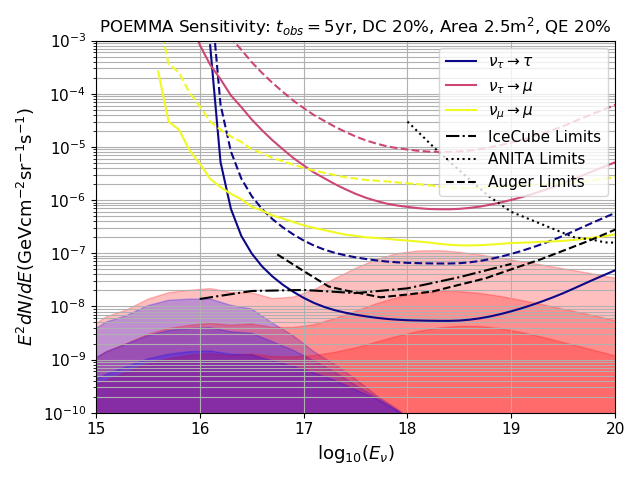}
\includegraphics[width=\linewidth]{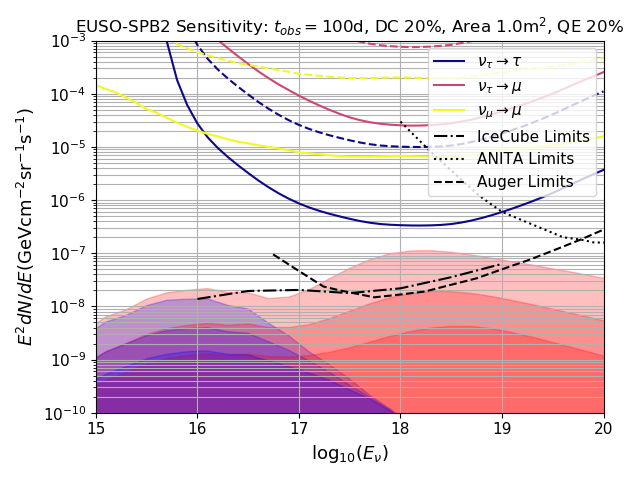}
\caption{Neutrino sensitivity scaled by neutrino energy squared for POEMMA [upper panel] and EUSO-SPB2 [lower panel], assuming duty cycle $20\%$ and flight times of 5~y and 100~d, respectively. Solid lines correspond to $\Delta\phi_E=360^{\circ}$ azimuthal field of view while dashed lines correspond to $\Delta\phi_E=30^{\circ}$ (POEMMA) and $\Delta\phi_E=12.8^{\circ}$ (EUSO-SPB2). The red and purple shaded regions represent the cosmogenic neutrino flux expected respectively in the case of a pure proton composition of UHECR and in the case of the mixed composition observed by Auger \cite{Aab:2019ogu}. Different neutrino fluxes correspond to different choices for the cosmological evolution of the UHECR sources as discussed in \cite{Aloisio:2015ega} (see text).}
\label{fig:sensitivities}
\end{figure}

As expected, at low neutrino energies ($<$ 10 PeV) EUSO-SPB2 is significantly more sensitive than POEMMA due to the near-field geometric effects for EAS observation from balloon altitudes that lead to a relative brightening of dim signals. At higher energies, the aperture of EUSO-SPB2 is roughly $\sim 1/5$ that of POEMMA, even under the assumption that both have $2\pi$ azimuth ranges (the ratio becomes $\sim 1/15$ taking into account the proper azimuth ranges). Additionally, we note the effect of muon induced cascades which, at low energies, is the dominant signal that allows for muon neutrino detection. At high energies, the muon neutrino signal becomes less significant with an acceptance that becomes roughly $\sim 1/5$ that of the tau neutrino curve. The least important detection channel for both instruments is that of secondary muons coming from \taon decay. While the showers from secondary muons can be equally as bright as those from primary muons, for low neutrino energies, they are not boosted by the improvement in \taon emergence probability and, at high neutrino energies, the decay length of the \taon limits the number of interactions the muon can experience before departing the atmosphere.

The only detection channel that can be directly compared with our results is that of the  \taon decay into hadrons and electrons, which has also been calculated in \cite{Reno:2019jtr} where the Cherenkov spatial distribution is fitted with a modified Gaussian profile in angle. Using the same Gaussian fit in angle to describe the Cherenkov photon spatial distribution, our curves are largely in agreement with \cite{Reno:2019jtr}. However, if we use the fit described in the text above, which models better the tails of the spatial profile for high altitude shower development, our calculated geometric aperture is larger than the geometric aperture computed in \cite{Reno:2019jtr} at neutrino energies greater  $10^{18}$~eV, with nearly an order of magnitude increase at $10^{20}$~eV.

The detector sensitivity to neutrino fluxes is defined as the minimum flux detectable by the instrument and, as discussed in \cite{Reno:2019jtr}, it can be computed as:

\begin{equation}
F_{\mathrm{sens}}(E_{\nu}) = \frac{2.44 N_{\nu}}{\mathrm{ln}(10)E_{\nu}\braket{A\Omega} (E_{\nu})t_{\mathrm{obs}}}
\label{sensi}
\end{equation}
where $N_{\nu}=3$ is the number of neutrino flavours and the numerical factor $2.44$ is the number of neutrino events (above threshold) required to be detected per decade in energy to reach a confidence level larger than $90\%$ \cite{Feldman_1998,Tanabashi:2018oca}. The total observation time $t_{\mathrm{obs}}$ is estimated to be 5~yr for the POEMMA mission and 100~d for the EUSO-SPB2 mission, both with similar duty cycles of $20\%$. 

The sensitivity curves for POEMMA and EUSO-SPB2 in the different neutrino detection channels (as labeled) are shown in figure \ref{fig:sensitivities}, with solid lines corresponding to an azimuth range of $2\pi$ and the dashed lines to the actual azimuth ranges of POEMMA and EUSO-SPB2. The shaded areas in figure \ref{fig:sensitivities} correspond to the cosmogenic neutrino flux, as computed in \cite{Aloisio:2015ega}, produced by the interaction of UHECR with astrophysical backgrounds. 

\begin{figure}[t!]
\includegraphics[width=\linewidth]{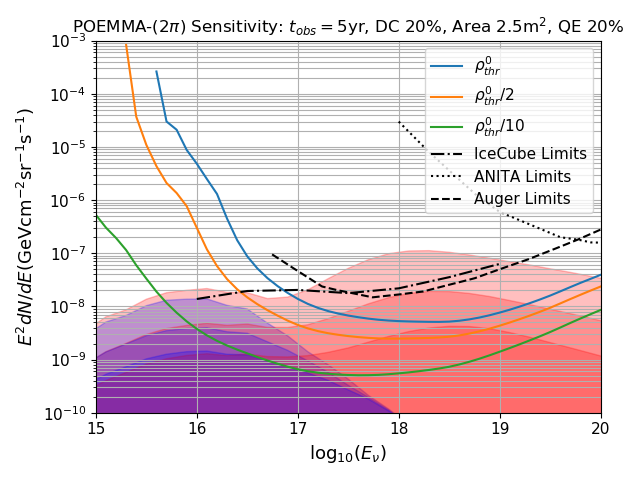}
\caption{Total sensitivity, summed over all detection channels shown in figure \ref{fig:sensitivities}, of a POEMMA like experiment with $2\pi$ azimuth range and different choices for $\rho_{thr}$. The solid blue line is the nominal case of the POEMMA design $\rho_{thr}=\rho^0_{thr}$, while the solid orange line corresponds to the case $\rho_{thr}=\rho^0_{thr}/2$ and the solid green line  corresponds to the case $\rho_{thr}=\rho^0_{thr}/10$.}
\label{fig:sensitivities2}
\end{figure}

The expected flux of cosmogenic neutrinos strongly depends on the UHECR mass composition and on the cosmological evolution of UHECR sources. In figure \ref{fig:sensitivities}, red shaded areas correspond to a pure proton composition of UHECR while the purple areas correspond to the mixed mass composition observed by Auger \cite{Aab:2019ogu}. Different choices of the cosmological evolution of UHECR sources are also plotted in figure \ref{fig:sensitivities} with the (shaded) upper curve corresponding to the cosmological evolution of Active Galactic Nuclei (AGN), the middle curve to the evolution of the Star Formation Rate (SFR) and the lower curve to the case of no cosmological evolution (see \cite{Aloisio:2015ega} and references therein).

As follows from figure \ref{fig:sensitivities}, the EUSO-SPB2 and POEMMA instruments are not suitable to detect the cosmogenic neutrino flux, with sensitivity curves comparable to current on-ground neutrino observatories only at the highest energies. On the other hand, the POEMMA instrument with an azimuth range of $2\pi$ shows a better sensitivity (by roughly one order of magnitude) with respect to on-ground observatories, even if it stil remains marginally sensitive only to the largest cosmogenic neutrino flux expected in the unlikely possibility of a pure proton composition of UHECR.

In more general terms, the sensitivity of a POEMMA-like detector is mainly dependent on the signal detection capabilities of the instrument which are fixed by the minimum density of Cherenkov photons (the threshold photon density $\rho_{thr}$ in equation \ref{delta}, given in photons/$m^{2}$) that can trigger the signal above the background. In figure \ref{fig:sensitivities2}, we plot the total sensitivity curves, i.e. summed over all detection channels, for a POEMMA-like instrument, with $2\pi$ azimuth range, corresponding to three different values for $\rho_{thr}$: the nominal case of POEMMA $\rho_{thr}=\rho^0_{thr}$ (blue solid line), the case with $\rho_{thr}=\rho^0_{thr}/2$ (orange solid line) and the extremely optimistic case with one order of magnitude improvement in detecting faint signals $\rho_{thr}=\rho^0_{thr}/10$  (green solid line), where $\rho^0_{thr}$ is again $20 \gamma/ m^{2}$ . As expected, reducing $\rho_{thr}$, the sensitivity of the detector increases more at low energies. This is due to the increased probability of detecting dim showers not visible if larger $\rho_{thr}$ are required to trigger on the signal. Note that this $\rho_{thr}$ scaling does not directly consider the effects of the dark-sky background on the detectability of the Cherenkov signal, but demonstrates the potential increase in sensitivity if the effects of the background can be reduced. 

The expected number of recorded events is given by:

\begin{equation}
N = \int \int \langle A\Omega \rangle (E_{\nu}) \Phi (E_{\nu}) d E_{\nu} dt
\label{Nevents}
\end{equation}
where $\Phi (E_{\nu})$ is the flux of cosmogenic neutrinos, integrated over the lifetime of the experiment and taking into account its duty cycle. 

In the tables below, we give the expected number of detected neutrino events in the energy range (1~PeV,100~EeV), for a POEMMA instrument, with $2\pi$ azimuth aperture, five years operating time and 20$\%$ duty cycle, in the different channels of detection and for the three different choices of $\rho_{thr}$ of figure \ref{fig:sensitivities2}. The first table refers to the case of a pure proton composition of UHECR while the second to the mixed composition as observed by Auger. For each UHECR composition model, we use the most optimistic cosmogenic flux model which has not yet been ruled out by current experiments. That is, for the pure proton composition model: SFR evolution, and for the mixed composition model: AGN evolution. The neutrino flux models for the various cosmological evolutions are taken from \cite{Aloisio:2015ega}.

\begin{table}[h!]
\begin{center}
\begin{tabular}{ |c|c|c|c| } 
 \hline
 & $\nu_{\tau} \rightarrow \tau$ & $\nu_{\tau} \rightarrow \mu$ & $\nu_{\mu} \rightarrow \mu$ \\
 \hline
$\rho^0_{thr}$ & 7.06 & $5.23 \times 10^{-2}$ & $3.05 \times 10^{-1}$ \\ 
\hline
$\rho^0_{thr}/2$ & 14.46 & $1.20 \times 10^{-1}$ & $6.62 \times 10^{-1}$ \\
\hline
$\rho^0_{thr}/10$ & 61.59 & $6.83 \times 10^{-1}$ & 3.51 \\
\hline
\end{tabular}
\caption{Maximum integrated cosmogenic neutrino events in the different detection channels for a POEMMA like mission with $2\pi$ azimuth range for different values of the minimum detectable photon density: the POEMMA case $\rho_{thr}=\rho^0_{thr}$, $\rho_{thr}=\rho^0_{thr}/2$ and $\rho_{thr}=\rho^0_{thr}/10$. A pure proton composition with SFR cosmological evolution of sources is assumed. The corresponding cosmogenic neutrino fluxes are taken from \cite{Aloisio:2015ega}.}
\label{table:protons}
\end{center}
\end{table}

\begin{table}[h!]
\begin{center}
\begin{tabular}{ |c|c|c|c| } 
 \hline
 & $\nu_{\tau} \rightarrow \tau$ & $\nu_{\tau} \rightarrow \mu$ & $\nu_{\mu} \rightarrow \mu$ \\
 \hline
$\rho^0_{thr}$ & $2.07 \times 10^{-1}$ & $8.12 \times 10^{-4}$ & $1.54 \times 10^{-2}$ \\ 
\hline
$\rho^0_{thr}/2$ & $7.99 \times 10^{-1}$ & $2.82 \times 10^{-3}$ & $6.31 \times 10^{-2}$ \\
\hline
$\rho^0_{thr}/10$ & 8.86 & $3.28 \times 10^{-2}$ & 1.03 \\
\hline
\end{tabular}
\caption{As in table \ref{table:protons} assuming the mixed mass composition of UHECR observed by Auger with AGN cosmological evolution.}
\label{table:nuclei}
\end{center}
\end{table}



\section{Conclusions}
\label{conclusions}

In this paper, we have presented a detailed computation of the optical Cherenkov signals produced by upward-moving EAS induced by \taons and muons. These phenomena are proxies for the interaction of high-energy (muon and tau) neutrinos inside the Earth and can be used to detect astrophysical neutrinos with energies larger than a few PeV. In our discussion we have focused the attention mainly on the detection capabilities of the POEMMA \cite{Anchordoqui:2019omw,Olinto:2019mjh,Olinto:2017xbi} and EUSO-SPB2 instruments \cite{Adams:2017fjh,Wiencke:2019vke}.

Building on the methodology of \cite{Reno:2019jtr}, we have improved the computation scheme in different ways: i) we used a more detailed evaluation of the neutrino propagation inside the Earth, and we included, for the first time, ii) the effects of muon neutrino induced showers and iii) the effects of the electron lateral spreading in the shower development. 

As discussed in section \ref{earth}, to model the neutrino propagation through the Earth, we used the NuTauSim Monte Carlo code \cite{Alvarez-Muniz:2017mpk}, which properly takes into account neutrino interactions, energy losses, and regeneration to provide as output an accurate evaluation of the lepton flux for air shower modeling. From this, we include also the sampling of shower energy from \taon decay, rather than using the average value. We also include, for the first time, the effects of muon induced showers from both \taon decay and muon neutrino interactions in the Earth. To do so, we appropriately modified NuTauSim to propagate muon neutrinos inside the Earth, and model the muon interactions in the atmosphere taking into account all relevant channels of energy losses (bremsstrahlung, pair production, photo-nuclear interaction) \cite{Alvarez-Muniz:2021mpk}. 

Including muon detection improves the sensitivity of balloon and low Earth orbit based instruments to the optical Cherenkov signal from upward going EAS, especially at low energies. This follows from two main facts: 1) muons have lifetimes significantly larger than $\tau$-leptons, thus boosting their Earth emergence probabilities at low energies and 2) muons, regardless of their initial energy, may begin showering high in the atmosphere, thus avoiding much of the optical atmospheric extinction, as well as reducing the distance to the instrument, both of which lead to significant brightening of the signal compared to \taons of comparable energy.

As discussed in sections \ref{upward} and \ref{rates}, we have carefully studied the analytical model to fit the Cherenkov photons distribution using an accurate approximation at large angles off shower axis. This regime is particularly relevant as it fixes the actual detection capabilities of the instrument. 

Our results for the sensitivity of POEMMA agree well with those in \cite{Reno:2019jtr} at low energies. With the inclusion of the muon induced cascades, we obtain an improved low energy sensitivity. At high energies, we obtain a better sensitivity primarily because of our improvements in modeling the extended tails of the Cherenkov spatial distribution. 

We have also discussed the sensitivity of the EUSO-SPB2 detector, in this case our detailed study of the electron lateral spreading was instrumental to model the near-field effects. Moreover, the inclusion of muon induced showers demonstrated that sub-orbital detectors can, in principle, test neutrino energies starting from the PeV regime, even if with a much reduced sensitivity.

Considering as reference value the cosmogenic neutrino flux computed in \cite{Aloisio:2015ega}, we proved that EUSO-SPB2 is not suitable to detect such particles, with a worse sensitivity than present and planned neutrino observatories on ground, and the POEMMA detector could perform better than on-ground observatories only in the ideal case of a $2\pi$ azimuth range. In the case of the actual POEMMA design, with an azimuth range of $30^{\circ}$, the detection capabilities are sensibly reduced and the diffuse neutrino flux cannot be observed, based on the current cosmogenic neutrino flux models. On the other hand, as recently pointed out in \cite{Venters:2019xwi}, the POEMMA satellite has an unprecedented capability to follow-up transient sources and with its sensitivity, can detect several possible bursting neutrinos sources such as those produced by black holes or neutron stars mergers \cite{Venters:2019xwi}. 

Finally, in order to bracket the detection capabilities of a generic low Earth orbit detector, with an ideal $2\pi$ azimuth range, we have considered a POEMMA-like experiment with different choices of the minimum detectable Cherenkov photon signal. It follows that to be sensitive to the cosmogenic neutrino flux, a low Earth orbit instrument should improve the detection capabilities reducing the minimum detectable Cherenkov signal by one order of magnitude, with respect to the current POEMMA design. This definitively represents a severe challenge from the experimental point of view, in light of the detrimental effects of the strong dark-sky background over the optical Cherenkov wavelength band (300~nm-1000~nm). 

\begin{acknowledgments}

We would like to thank those who were instrumental in discussing the physics and computations required for this work, including M. Bertaina, F. Bisconti, P. Blasi, I. De Mitri, J. Eser, D. Fargion, F. Fenu, S. Grimm, D. Grisham, S. Petrera,  M. Hall Reno, F. Salamida, Tonia Venters, and D. Wright, as well as the POEMMA and EUSO-SPB2 collaborations. For the use of CORSIKA in this work, we would also like to thank D. Heck and T. Pierog. Additionally, we would like to thank P. Ilten regarding his continued help with PYTHIA.
\end{acknowledgments}

\bibliographystyle{apsrev4-1}
\bibliography{biblio+JFK.bib} 

\end{document}